\documentclass[]{iscram}

\usepackage[utf8]{inputenc}

\usepackage{lipsum}
\usepackage{tikz}
\usepackage{fancyvrb}
\usepackage{listings}
\usepackage{enumitem}
\usepackage{tcolorbox}
\usepackage{multicol}
\usepackage{siunitx}
\usepackage{xpatch}
\usepackage{svg}
\usepackage{multirow}
\usepackage{lscape}
\usepackage{appendix}

\lstdefinestyle{common}{
  xleftmargin=.5em,
  xrightmargin=.5em,
  frame=single,framesep=.5em,framerule=0pt,
  fancyvrb=true,
  basicstyle=\ttfamily,
  keywordstyle=\color{cyan!50!blue!75!black}\bfseries,
  commentstyle=\color{red!50!black}\itshape,
  stringstyle=\ttfamily\color{green!50!black},
  numbers=none,
  showspaces=false,
  showstringspaces=false,
  fontadjust=true,
  keepspaces=true,
  flexiblecolumns=true,
  emphstyle=\color{red},
}
\lstdefinestyle{TeX}{
  style=common,
  backgroundcolor=\color{blue!5},
  aboveskip=5pt,
  belowskip=5pt,
  language=[LaTeX]TeX,
  moretexcs={
    abstract, addbibresource, iscramset, keywords, mainmatter,
    maketitle, printbibliography, subsection, subsubsection, url,
    urldef, href, includegraphics, ldots, parencite, citeauthor,
    citeyear, citetitle, midrule, toprule, bottomrule
  },
  fancyvrb=true,
}
\lstdefinestyle{console}{
  style=common,
  backgroundcolor=\color{gray!10},
  aboveskip=5pt,
  belowskip=5pt,
}

\newlist{options}{description}{1}
\setlist[options]{%
  beginpenalty=10000,%
  itemsep=.5\parskip plus .3\parskip minus .2\parskip,
  parsep=.5\parskip plus .3\parskip minus .2\parskip,
  topsep=.5\parskip plus .3\parskip minus .2\parskip,
  partopsep=.5\parskip plus .3\parskip minus .2\parskip,
  style=nextline,labelindent=1em,%
  font=\normalfont\ttfamily}

\colorlet{macro color}{cyan!50!blue!75!black}
\colorlet{option color}{red!50!black}
\colorlet{generic color}{green!40!black}

\newtcolorbox{pseudoTeX}{colback=blue!5,colframe=blue!5,before=\nobreak}
\let\LaTeXorig\LaTeX
\renewcommand\LaTeX{\bgroup\fontfamily{lmr}\selectfont\upshape\LaTeXorig\egroup}

\urldef{\sitewebgind}\url{http://gind.mines-albi.fr}
\urldef{\sitewebminesalbi}\url{http://www.mines-albi.fr}
\urldef{\gmailpaulgaborit}\url{email adress}
\urldef{\jdmail}\url{...@example.com}
\urldef{\sitex}\url{www.example.com}

\iscramset{
  CoRe Paper 2023={Social Media for Crisis Management},
  title={A Twitter narrative of the COVID-19 pandemic in Australia},
  short title={A Twitter narrative of the COVID-19 pandemic in Australia},
  author={
    short name={Lamsal},
    full name=Rabindra Lamsal\thanks{corresponding author},
    affiliation={
      The University of Melbourne\\
       \href{mailto:rlamsal@student.unimelb.edu.au}{rlamsal@student.unimelb.edu.au}
    },
  },
  author={
    full name= Maria Rodriguez Read,
    affiliation={
      The University of Melbourne\\
       \href{mailto:maria.read@unimelb.edu.au}{maria.read@unimelb.edu.au}
    },
  },
  author={
    full name= Shanika Karunasekera,
    affiliation={
      The University of Melbourne\\
       \href{mailto:karus@unimelb.edu.au}{karus@unimelb.edu.au}
    },
  },
}

\usepackage{tikzpagenodes}
\usetikzlibrary{fit}

\begin{document}

\maketitle


\abstract{
Social media platforms contain abundant data that can provide comprehensive knowledge of historical and real-time events. During crisis events, the use of social media peaks, as people discuss what they have seen, heard, or felt. Previous studies confirm the usefulness of such socially generated discussions for the public, first responders, and decision-makers to gain a better understanding of events as they unfold at the ground level. This study performs an extensive analysis of COVID-19-related Twitter discussions generated in Australia between January 2020, and October 2022. We explore the Australian Twitterverse by employing state-of-the-art approaches from both supervised and unsupervised domains to perform network analysis, topic modeling, sentiment analysis, and causality analysis. As the presented results provide a comprehensive understanding of the Australian Twitterverse during the COVID-19 pandemic, this study aims to explore the discussion dynamics to aid the development of future automated information systems for epidemic/pandemic management. 
}


\section{Introduction}
The gravity of the COVID-19 pandemic made people more vocal on social media platforms, especially on microblogs such as Twitter. Twitter discussions, i.e. tweets, specific to the pandemic, collected by researchers and laboratories globally, have been reported to be in billions (\cite{chen2020tracking,781w-ef42-20,imran2022tbcov,lamsal2023billioncov}), with the United States, United Kingdom, Canada, and India generating the most discussions for both English-only and multilingual  discourse. The Australian Twitterverse also seemed significantly vocal towards the pandemic; (\cite{781w-ef42-20,imran2022tbcov}) report Australia's presence within the top 10 countries generating the most discussions on Twitter regarding the pandemic. The pandemic introduced numerous events within Australia in 2020--22~---~notable ones include over a hundred film and TV show productions getting halted, official interest rates cut to a record low, the introduction of economic stimulus packages from federal and state governments, a recession for the first time in nearly three decades, agriculture workers shortage, local council elections getting disrupted, protests against lockdown restrictions and the national vaccine program, and the suspension of sporting events. Discussions related to these events and numerous global topics specific to the pandemic have been trending on the Australian Twitterverse since the COVID-19 outbreak.

\subsection{Social media and situational awareness}
The main reasons for the growth of social media are speed, transparency, and ubiquity, assisted by rapid developments in mobile technology. The events that would have remained obscure for a long course are now being reported and shared worldwide within seconds (\cite{mayfield2011commander}). Social media platforms provide an edge over traditional ways, such as individual cellular communications, by providing a public broadcast platform for individuals. Such broadcast platforms help engage people with the exchange of status updates, stories, and media items, which have been reported to have a significant advantage for extracting ``situational awareness" (\cite{imran2015processing,lamsal2022socially}). During mass emergencies, compared to normal hours, people tend to make use of social media excessively to get situational updates (\cite{vieweg2012situational}) and such socially generated discussions accumulate to hundreds of thousands and even millions in cases such as the COVID-19 pandemic (\cite{lamsal2022socially}). Those discussions, if timely monitored, processed, and analyzed, can contain actionable information relating to the event (\cite{hughes2009twitter,vieweg2010microblogging,vieweg2012situational}) that can assist first responders and decision-makers to come up with efficient plans for effective disaster management. Literature in the crisis computing discipline (\cite{lamsal2022socially,imran2015processing}) summarizes the effectiveness of social media discussions for numerous humanitarian aid-related tasks, with ``situational awareness'' as one of the most evident. 

This paper presents an extensive retrospective Twitter narrative of the Australian Twitterverse during the COVID-19 pandemic by employing state-of-the-art supervised and unsupervised machine learning approaches. We perform (i) network analysis on hashtags and mentions, (ii) topic modeling with neural embeddings, (ii) sentiment analysis with a transformer-based language model, and (iv) causal analysis with Granger causality tests. Through the study of hashtags and mentions networks, we seek to distinguish the significance of country-level and state-specific hashtags and identify the classifications of Twitter accounts that generate the most engagement during a pandemic. With topic modeling and sentiment analysis, we aim to extract topics (events) that generate the highest tweet interest and investigate the sentiment trends during different pandemic phases. And with causal analysis, we seek to study the causality behavior of discussion-based time series on the confirmed cases and death cases time series. These analyses, combinedly, contribute to the during-disaster and post-disaster phases of disaster management, and the results assist in understanding the conversational dynamics of a pandemic to aid the development of future information systems for epidemic/pandemic management.

\textbf{Contributions}: As per our knowledge, this study, with a timeline of almost 145 weeks (January 2020 to October 2022), provides the most comprehensive social media narrative of the COVID-19 pandemic in Australia through multiple analyses. In doing so, presented results aid in identifying areas with room for improvements for designing robust automated information systems for epidemic/pandemic management. Further, we release a large-scale geotagged tweets dataset\footnote{\texttt{https://dx.doi.org/10.21227/42h1-ge40}}, curated as a part of this study using the \textit{Full-archive search endpoint}\footnote{https://developer.twitter.com/en/docs/twitter-api/tweets/search/quick-start/full-archive-search} (this endpoint returns the entire volume of historical tweets).

\section{Literature Review}
Information systems researchers and practitioners have been formulating frameworks and tools to monitor, collect, analyze, summarize, and visualize social media data to assist in making timely and effective decisions during crisis events (\cite{imran2015processing}). The current literature heavily relies on microblog platforms, especially Twitter and Weibo, as the primary data source for designing such frameworks and tools~---~these platforms have large active user bases, their contents carry real-time attributes, and they provide multiple API endpoints for easy access to their public feed. Previous studies have corroborated the relevance of Twitter discussions in the management and analysis of emergency situations during all three phases of a disaster (\cite{martinez2018twitter})~---~especially, designing emergency monitoring, event detection and decision support systems, identifying relevant contents, performing rapid assessments, and visualizing the spatial and temporal contexts of disasters. Refer to these studies (\cite{vieweg2012situational,imran2015processing,lamsal2022socially}) for a thorough review of the literature related to the use of social media data for ``situational awareness" and associated methods, data sets, and algorithms.

Researchers have been collecting and sharing large-scale Twitter datasets to enable further research in better understanding the COVID-19 discourse. Tweets in (\cite{chen2020tracking,781w-ef42-20,banda2021large,imran2022tbcov,lamsal2023billioncov}) are in hundreds of millions and these datasets are some of the largest COVID-19-specific tweets collections at present. These datasets are based on \textit{streaming endpoint} whose payload returns 1\% of the entire Twitter data at a particular time. The dataset used in this study is based on \textit{Full-archive search endpoint}, which returns the entire volume of historical tweets. Our dataset also complements \textit{MegaGeoCOV} (\cite{lamsal2022twitter}) with the use of additional keywords and an extended collection period.

Topic modeling has been rigorously performed to identify latent topics across multiple thematic areas of Twitter discourse, including public health (\cite{ghosh2013we}), sporting events (\cite{steinskog2017twitter}), and disasters (\cite{alam2018twitter}). Multiple studies (\cite{boon2020public,xue2020public,abd2020top,lamsal2022twitter}) have performed topic modeling on COVID-19-specific Twitter discussions to explore the public perception of the pandemic and uncover the trends and themes of concerns tweeted by individuals. The majority of the existing studies use \textit{bag-of-words}-based techniques (\cite{lamsal2022socially}), which consider documents as bag-of-words and model individual documents as a mixture of latent topics. The bag-of-words representation fails to capture the true semantics of words, leading to a possibly imprecise representation of documents. Recent progress in natural language processing has come through the use of \textit{contextual embeddings}, such as from ELMo, BERT, and GPT-3, which handle the issues associated with semantic similarity and polysemous. As a result, neural embeddings have been used in topic modeling giving rise to neural topic models (\cite{angelov2020top2vec,zhao2021topic,grootendorst2022bertopic}). This study employs neural embeddings-based topic modeling.

Sentiment analysis is one of the most explored research fields for analyzing people's opinions, and attitudes toward factors such as situations, individuals, products, and organizations. Sentiment analysis techniques, in general, are machine learning-based, lexicon-based, and hybrid (\cite{medhat2014sentiment}). More recently, transformer-based (\cite{vaswani2017attention}) models, such as BERT, RoBERTa, and XLNet, are being used for designing high-performing sentiment analyzers. In the case of tweets, BERTweet (\cite{nguyen2020bertweet}), a RoBERTa-based (\cite{liu2019roberta}) language model pre-trained on millions of tweets, seems to produce state-of-the-art results in part-of-speech tagging, named-entity recognition, and text classification tasks.

Granger causality (\cite{Granger}) analysis provides a powerful approach for performing causal inference, by testing whether a time series helps forecast another time series. The current literature employs Granger causality tests on time series data across fields such as neuroscience (\cite{Seth3293}), tourism and economy (\cite{NikolaosTourism,AkinboadeTourism}), stock markets (\cite{BOLLEN20111}), and foreign direct investments (\cite{HoffmannFDI}). Twitter discussions have also been studied for their causal behavior towards stock markets (\cite{BOLLEN20111}), cryptocurrencies (\cite{SHEN2019118}), elections (\cite{BovetFake}), and public health (\cite{lamsal2022twitter}). In this study, we perform Granger causality tests on time series data generated during topic modeling and sentiment analysis.

For network analysis of social media data~---~exploring and understanding graphs formed by socially generated data, such as tweets~---~tools such as Networkx, Gephi, Pajek, and IGraph are available (\cite{6821424}). Twitter discussion-based networks are majorly studied to develop an understanding of online communities (\cite{Cheong2011,Mena2015,Loni,Ahmed5G,lamsal2021design}). In this study, we perform network analysis of [state$\rightarrow$hashtag] and [state$\rightarrow$mention] relationships to identify state-specific concerns and highly engaging Twitter accounts, and we use Gephi (\cite{ICWSM09154}) for graph-based visual analytics.

\section{Overview of the pandemic in Australia}

Australia, one of the few countries around the world to adopt the zero-covid ``suppression with a goal of no community transmission'' public health policy during the COVID-19 pandemic, implemented controls on international travel and response to local outbreaks with stringent lockdowns and thorough contact tracing of local COVID-19 clusters. The country closed its international borders to the outside world for almost two years while imposing strict limits on local movements across its states and territories, thus often being referred to as ``Fortress Australia''.

Australia's mitigation strategies included early interventions to international travel to reduce transmissions from other countries, suppressing the growth of local COVID-19 clusters with exhaustive contact tracing, early recruitment of contact tracing workers, and use of intense lockdowns. The country had its first public COVID-19 vaccination on February 21, 2021, and by the early last quarter of 2021, 80\% of the eligible population (i.e. age$\ge$16) was administered at least a single dose of the COVID-19 vaccine. By the end of March 2022, this percentage increased to 95.0\%.
The country opened its borders on February 21, 2022, for all fully vaccinated people, and further restrictions on international travel under the Biosecurity Act were lifted on April 18, 2022, thus effectively opening up the country to the world. Although the country's mitigation strategies were in contrast to the ones implemented by other countries and territories worldwide, compared to the United States, the United Kingdom, and European countries, the COVID-19 numbers in Australia have been significantly lower until 2022.

\section{Data collection}

\begin{table}[t]
\caption{Keywords and hashtags used for data collection.}
    \label{keywords-hashtags}
    \centering
    \begin{tabular}{p{15.5cm}}
    \hline
         coronavirus, \#coronavirus, covid, \#covid, covid19, \#covid19, covid-19, \#covid-19, corona, \#corona, sarscov2, \#sarscov2, sars cov2, sars cov 2, covid\_19, \#covid\_19, \#ncov, ncov, \#ncov2019, ncov2019, 2019-ncov, \#2019-ncov, \#2019ncov, 2019ncov, pandemic, \#pandemic, quarantine, \#quarantine, \#lockdown, lockdown, ppe, n95, \#ppe, \#n95, pneumonia, \#pneumonia, virus, \#virus, mask, \#mask, vaccine, vaccines, \#vaccine, \#vaccines, lungs, flu, flatten the curve, flattening the curve, \#flatteningthecurve, \#flattenthecurve, hand sanitizer, \#handsanitizer, social distancing, \#socialdistancing, work from home, \#workfromhome, working from home, \#workingfromhome, \#covidiots, covidiots, herd immunity, \#herdimmunity, chinese virus, \#chinesevirus, wuhan virus, \#wuhanvirus, kung flu, \#kungflu, wearamask, \#wearamask, wear a mask, corona vaccine, corona vaccines, \#coronavaccine, \#coronavaccines, face shield, \#faceshield, face shields, \#faceshields, health worker, \#healthworker, health workers, \#healthworkers, \#stayhomestaysafe, \#coronaupdate, \#frontlineheroes, \#coronawarriors, \#homeschool, \#homeschooling, \#hometasking, \#masks4all, \#wfh, wash ur hands, wash your hands, \#washurhands, \#washyourhands, \#stayathome, \#stayhome, \#selfisolating, self isolating \\
         \hline
    \end{tabular}
\end{table}

In this study, we used Twitter's Full-archive search endpoint to collect global COVID-19-specific geotagged tweets created between January 1, 2020, and October 9, 2022. More than 90 keywords and hashtags were used with \texttt{has:geo} and \texttt{lang:en} operators while querying the endpoint to collect tweets that are geotagged and written in English. The hashtags and keywords are listed in Table \ref{keywords-hashtags}. We collected 17,826,615 tweets originating from 245 countries and territories worldwide. These tweets are geotagged with either point location or place information. Since the geo attributes for retweets are NULL, the collected data do not include retweets.

Table \ref{con-city-sour} gives the distributions (top 10) of tweets based on country, city, and source. Since we collected only English tweets, the dominance of native English-speaking nations is evident. Tweet distribution across countries ranks Australia fifth for generating the most tweets. Melbourne, Victoria is ranked eighth in terms of cities. Twitter native apps for iPhone, Android, and iPad, Instagram, and dlvr.it are the top sources of geotagged tweets. We filtered Australian tweets from the global corpus by conditioning the \texttt{geo.country} tweet object. In total, 481,944 tweets in the corpus were identified as originating from Australia. 

\begin{figure}[t!]
    \centering
    \includegraphics[width=1\textwidth]{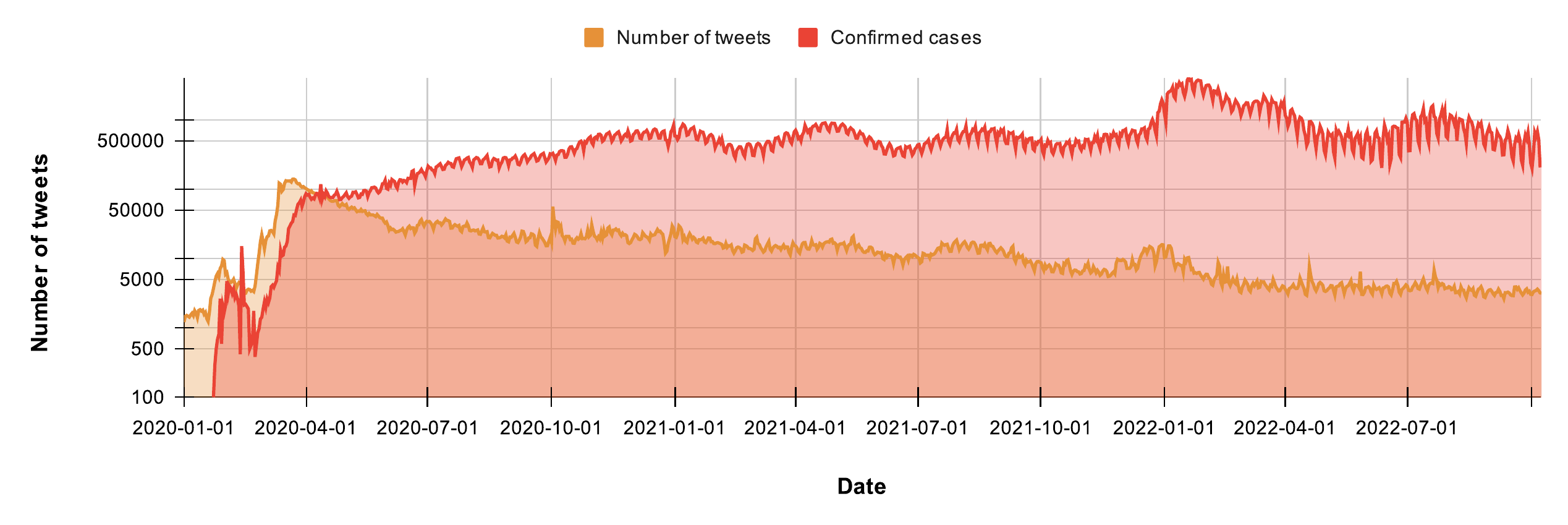}
    \caption{Daily distribution of collected tweets and confirmed cases (worldwide). $Y$-axis is in log scale.}
    \label{global-distribution}
\end{figure}

\begin{figure}[t!]
    \centering
    \includegraphics[width=1\textwidth]{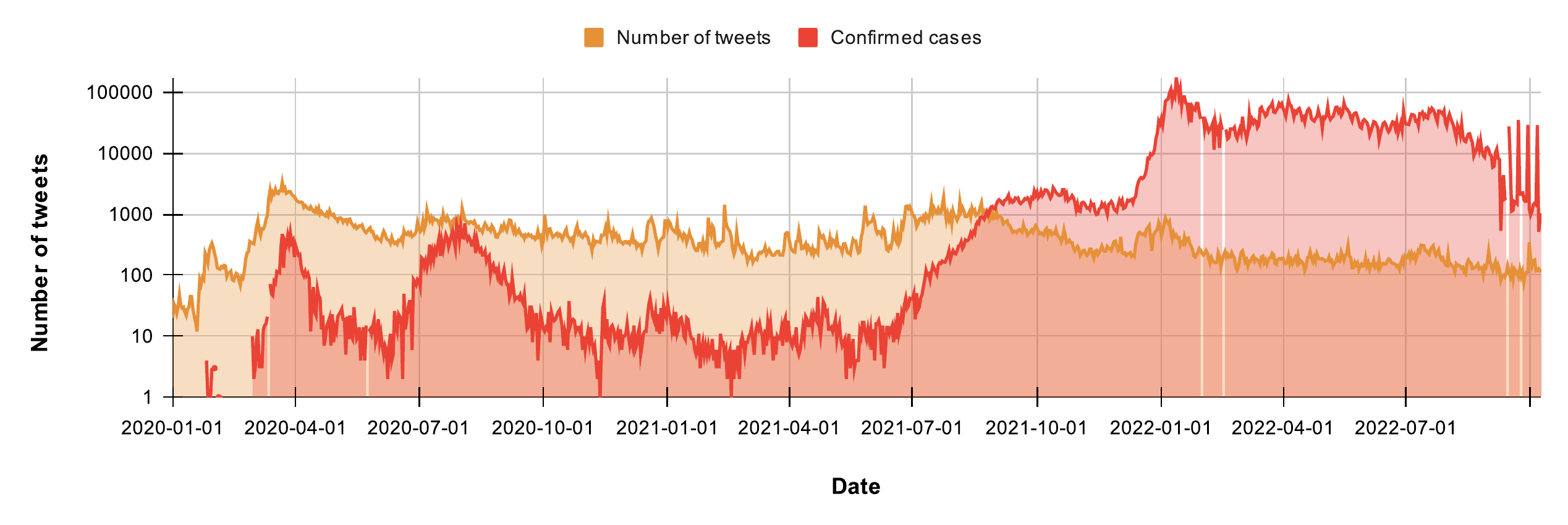}
    \caption{Daily distribution of collected tweets and confirmed cases (Australia). $Y$-axis is in log scale. Vertical white lines represent adjusted (in negative) values.}
    \label{aus-distribution}
\end{figure}

We performed some basic exploratory data analysis on the collected tweets: Figure \ref{global-distribution} and Figure \ref{aus-distribution} present the daily distribution of tweets alongside the daily confirmed cases in the world and Australia, respectively, Table \ref{top-aus-regions} lists most tagged Australian geolocations, and Figure \ref{geoplot-aus} is a geographical plot of the spatial distribution of tweets in Australia. The state-wise volume of tweets had the following order: Victoria, New South Wales, Queensland, Western Australia, Southern Australia, Australian Capital Territory, Tasmania, and Northern Territory. Melbourne (Victoria) and Sydney (New South Wales) seem to be participating in the discourse significantly compared to other major cities of Australia.

\begin{table}[t!]
\caption{Distributions of tweets based on country, city, and source (global). We list only the top 10 entries.}
\begin{minipage}[c]{0.30\textwidth}
    \centering
    \begin{tabular}{c|c}
    \hline
      \textbf{Country} & \textbf{tweets}\\
      \hline
      United States	& 8,792,388\\
      United Kingdom & 2,840,423\\
      India & 1,494,866\\
      Canada & 915,229\\
      Australia & 481,944\\
      South Africa & 406,316\\
      Nigeria & 336,515\\
      Ireland & 264,582\\
      Philippines & 200,404\\
      Pakistan & 116,879\\
      \hline
      \end{tabular}
    \end{minipage} 
    \hfill
  \begin{minipage}[c]{0.3\textwidth}
    \centering
    \begin{tabular}{c|c}
    \hline
      \textbf{City} & \textbf{tweets}\\
      \hline
      Los Angeles, CA & 286,393\\
      Manhattan, NY	& 218,571\\
      New Delhi, India	& 174,107\\
      Toronto, Ontario	& 171,569\\
      Mumbai, India	& 166,001\\
      Chicago, IL &	148,641\\
      Florida, USA (state)	& 144,245\\
      Melbourne, Victoria &	142,895\\
      Houston, TX & 133,062\\
      Brooklyn, NY & 132,873\\
      \hline
      \end{tabular}
  \end{minipage}
  \hfill
  \begin{minipage}[c]{0.3\textwidth}
    \centering
    \begin{tabular}{c|c}
    \hline
      \textbf{Source} & \textbf{tweets}\\
      \hline
Twitter for iPhone &	9,568,155\\
Twitter for Android &	6,791,426\\
Instagram &	830,623\\
Twitter for iPad &	313,718\\
dlvr.it &	81,657\\
Tweetbot for iOS &	62,372\\
Twitter Web App &	25,832\\
Twitter Web Client &	15,483\\
Hootsuite Inc. &	15,297\\
Twitter for Mac &	12,516\\
      \hline
      \end{tabular}
  \end{minipage}
  \label{con-city-sour}
\end{table}

\begin{minipage}{0.49\textwidth}
    \includegraphics[width=1\textwidth]{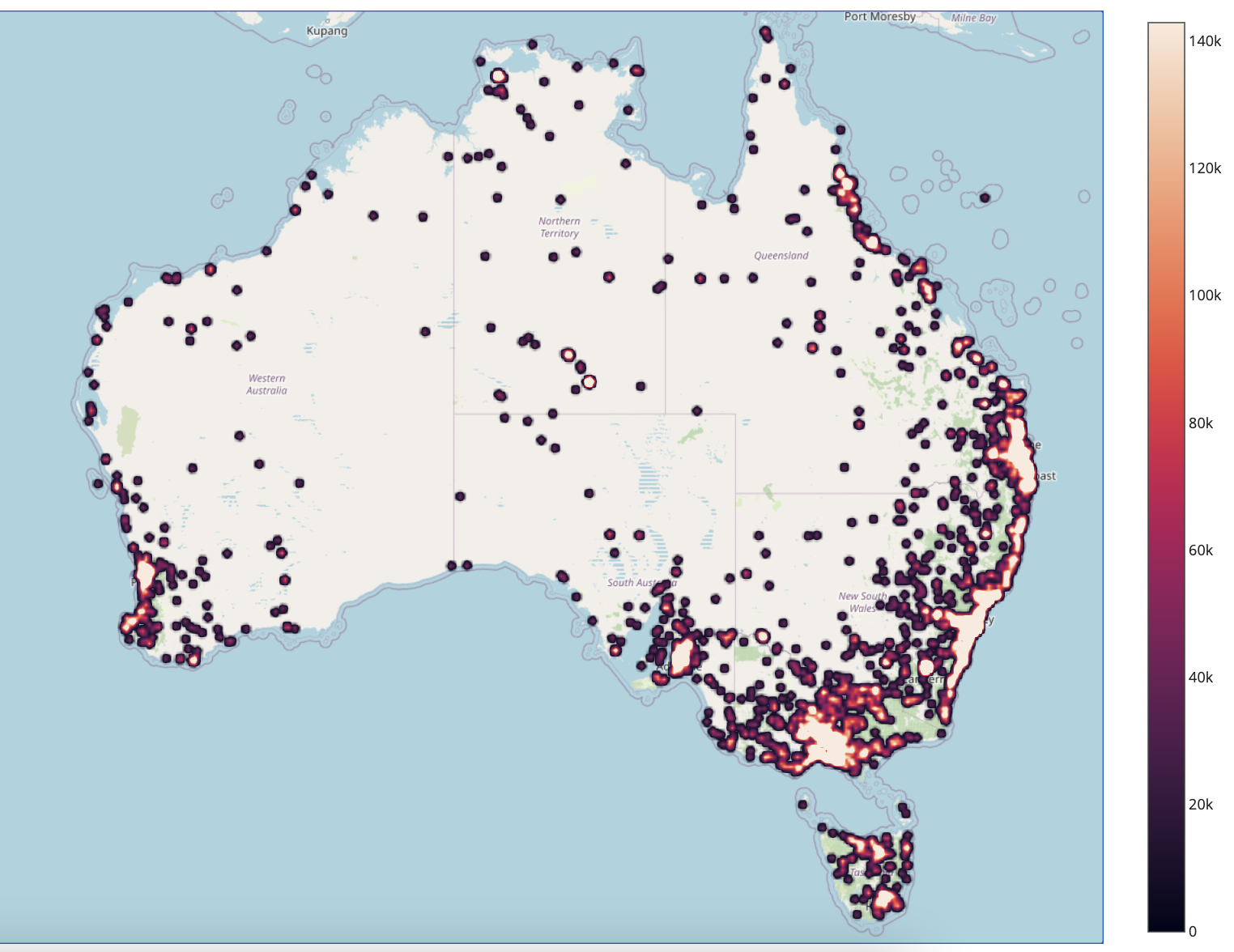}
    \captionof{figure}{Geoplot showing spatial distribution of COVID-19-specific tweets within Australia. Color scale represents number of tweets.}
  \end{minipage}
  \hfill
  \begin{minipage}{0.49\textwidth}
    \centering
    \captionof{table}{Top Australian regions in the COVID-19 discourse. Based on \texttt{geo.full\_name} tweet object.}
    \label{top-aus-regions}
    \label{geoplot-aus}
    \begin{tabular}{c|c}\hline
    \textbf{Place} & \textbf{tweets}\\
    \hline
      Melbourne, Victoria &	142,895\\
Sydney, New South Wales &	108,511\\
Brisbane, Queensland &	35,176\\
Perth, Western Australia &	29,588\\
Adelaide, South Australia &	21,990\\
Canberra, Australian Capital Territory &	12,283\\
Gold Coast, Queensland &	11,302\\
Victoria, Australia &	7,829\\
New South Wales, Australia &	6,583\\
Newcastle, New South Wales &	5,650\\
Sunshine Coast, Queensland &	4,280\\
Central Coast, New South Wales &	3,736\\
Tasmania, Australia &	3,635\\
Hobart, Tasmania &	3,469\\
\hline
      \end{tabular}
    \end{minipage}

\section{Experiments, Results, and Discussions}
\subsection{Data pre-processing}
We performed basic text pre-processing tasks on the collected tweets: replacing (i) URLs with the $<$HTTPURL$>$ token, (ii) HTML entities with their character representation, (iii) emojis with the $<$EMOJI$>$ token, and removing unnecessary spaces, indentations, and section breaks.

\subsection{Analyses}
\subsubsection{Exploring hashtags and mentions usage at the geo-level}
Hashtags~~---~~keywords or phrases prefaced by the hash sign (\#)~---~have always been a go-to method for people to categorize, search and join discussions related to a particular event. Since the outbreak, many COVID-19-specific keywords and phrases evolved and were in use while referencing the pandemic. The Twitterverse received hundreds of hashtags related to the pandemic, some notable and globe-specific ones include \#coronavirus, \#covid19, \#sarscov2, \#pandemic, \#quarantine, \#flattenthecurve, \#handsanitizer, \#workfromhome, \#herdimmunity, \#stayhomestaysafe, \#frontlineheroes, \#homeschooling, and \#faceshields, among others. Country/territory-specific hashtags were also in use. Therefore, with a network analysis, we seek to identify hashtags that were specifically used in each of the Australian states to discuss pandemic-related situations.

The [state$\rightarrow$hashtag] relationships form a dense block of nodes representing the hashtags common to multiple states, while state-specific hashtags are represented by sparsely connected blocks of nodes. Timely construction of similar \texttt{[county$\rightarrow$hashtag]} and \texttt{[district$\rightarrow$hashtag]} networks and mining of tweets based on these relationships can assist in understanding region-specific concerns. Similarly, we also perform network analysis of \texttt{[state$\rightarrow$mention]} relationships to identify the most notable Twitter accounts in the Australian COVID-19 discourse. Such highly-engaging accounts can assist in the timely dissemination of factual information and aid in fighting misinformation, to a significant extent.

\begin{figure}
\begin{minipage}[c]{0.49\textwidth}
        \centering
        \label{hash-out-degree}
        \includegraphics[width=0.8\textwidth]{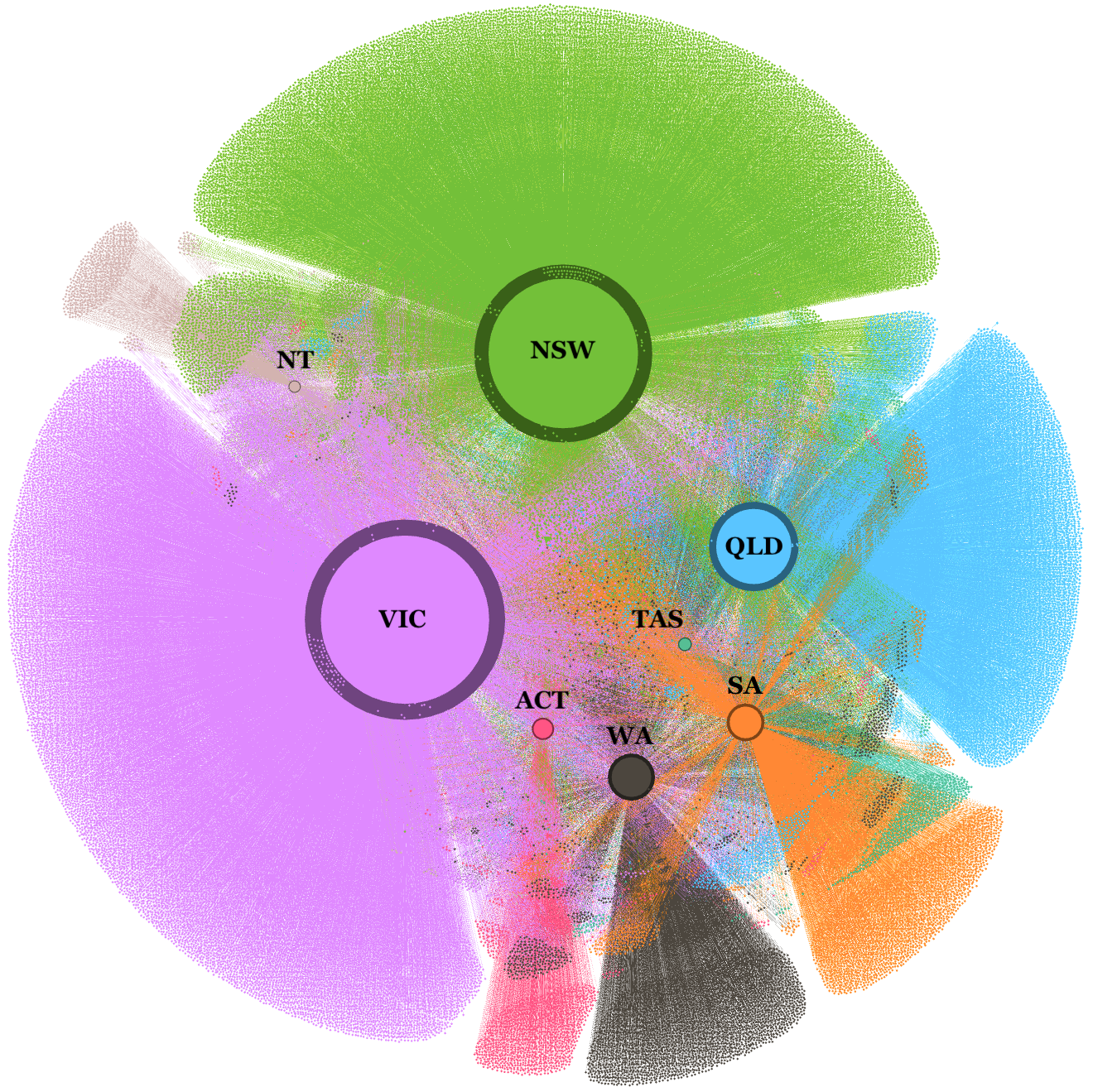}
        \caption*{(a) out-deg. (hashtags)}
        \label{hash-out-degree}
\end{minipage}
\begin{minipage}[c]{0.49\textwidth}
        \centering
        \includegraphics[width=0.8\textwidth]{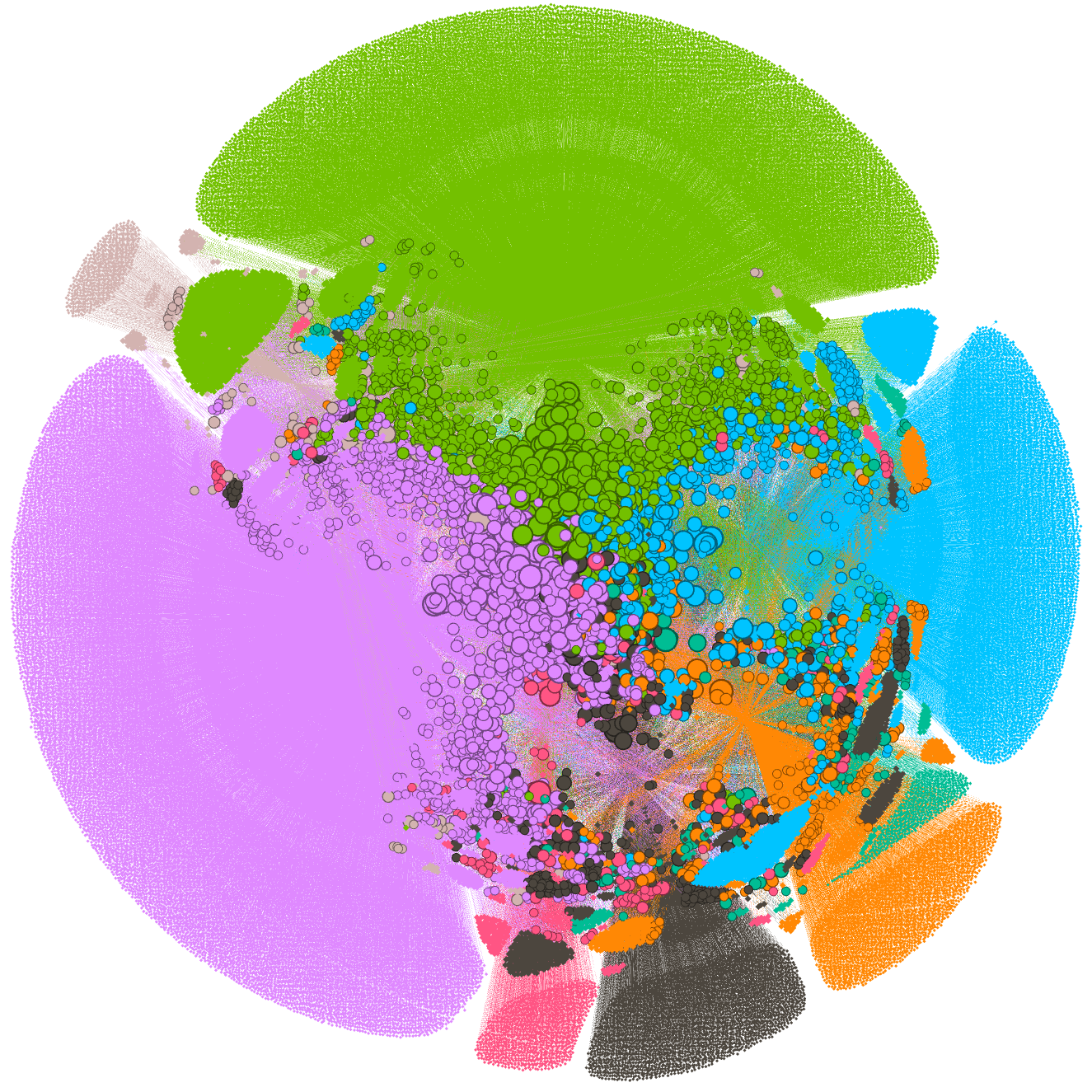}
        \caption*{(b) in-deg. (hashtags)}
        \label{hash-in-degree}
\end{minipage}

\begin{minipage}[c]{0.49\textwidth}
        \centering
        \label{hash-out-degree}
        \includegraphics[width=0.8\textwidth]{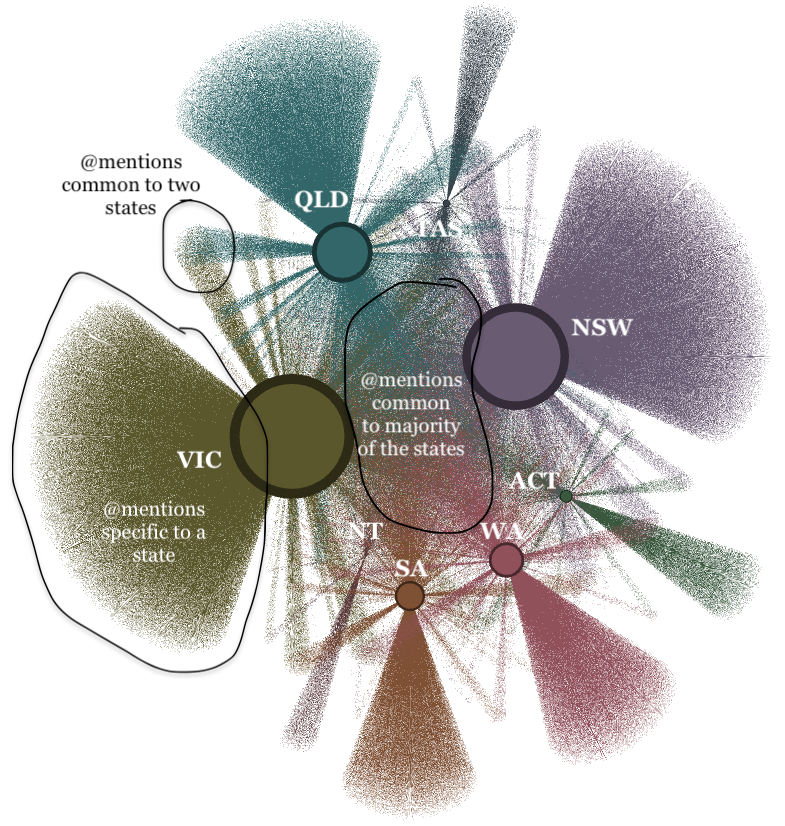}
        \caption*{(c) out-deg. (mentions)}
        \label{mention-out-degree}
\end{minipage}
\begin{minipage}[c]{0.49\textwidth}
        \centering
        \includegraphics[width=0.8\textwidth]{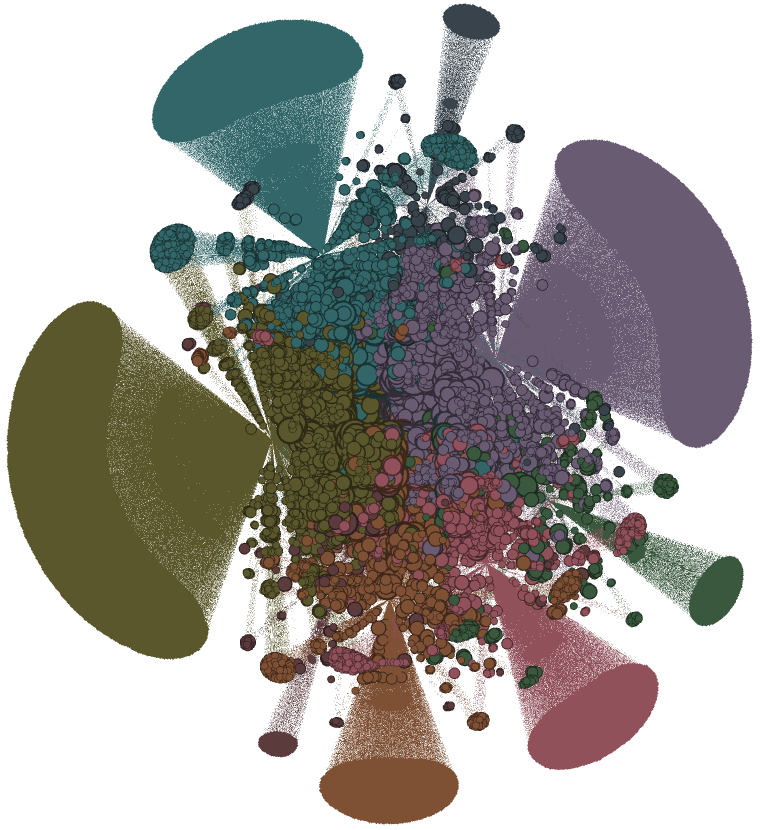}
        \caption*{(d) in-deg. (mentions)}
        \label{mention-in-degree}
\end{minipage}
\caption{Visualization of hashtags and mentions networks.}
\label{net-ana}
\end{figure}

\begin{table}
    \centering
    \caption{Out-degree information of Australian states in the hashtags and mentions networks.}
    \begin{tabular}{c|c|c|c|c}
    \hline
    & \multicolumn{2}{c|}{\textbf{\#hashtags network}} & \multicolumn{2}{c}{\textbf{@mentions network}}\\
    \hline
       \textbf{State}  & \textbf{out-degree} & \textbf{Wgt. out-degree} & \textbf{out-degree} & \textbf{Wgt. out-degree}\\
         \hline
       Victoria (\textit{VIC})  & 29,387 & 122,745 & 53,809 & 173,948\\
       New South Wales (\textit{NSW}) & 26,071 & 101,770 & 46,001  & 138,221\\
       Queensland (\textit{QLD}) & 12,912 & 41,335 & 26,343 & 78,578\\
       Western Australia (\textit{WA}) & 6,641 & 20,001 & 15,263 &  37,694\\
       South Australia (\textit{SA}) & 5,247 & 14,514 & 13,072  & 31,291\\
       Australian Capital Territory (\textit{ACT}) & 2,954 & 8,827 & 5,556 & 12,336\\
       Tasmania (\textit{TAS}) & 1,662 & 4,777 & 3,183 & 5,971\\
       Northern Territory (\textit{NT}) & 1,478 & 2922 & 1,751 & 3,307\\
         \hline
    \end{tabular}
    \label{outdegree-table}
\end{table}

We deployed a planet-level geocoding endpoint powered by OpenStreetMap data\footnote{https://www.openstreetmap.org/} to extract state information for each place name in \texttt{geo.full\_name} tweet object. This step was necessary to normalize place names such as ``Melbourne, Victoria" and ``Victoria, Australia". In total, 317,158 \texttt{[state$\rightarrow$hashtag]} relationships were generated for the hashtags network and 523,857 \texttt{[state$\rightarrow$mention]} relationships for the mentions network. The hashtags network had 65,022 nodes and 86,352 edges, while the mentions network had 123,220 nodes and 164,978 edges. The visual representations of the networks, in terms of out-degree and in-degree relationships, are shown in Figure \ref{net-ana}. State-specific out-degree information is provided in Table \ref{outdegree-table}. Results from the out-degree analysis show that Victoria used the highest number of (unique) hashtags and mentions, followed by New South Wales and Queensland. The in-degree analyses on hashtags and mentions are summarized in Table \ref{hashtags-all-states}, Table \ref{hashtags-single-states}, and Table \ref{most-mentions}. For a detailed outlook on the most common hashtags, Table \ref{hashtags-all-states} provides in-degree (suggesting the number of states that used the hashtag) and weighted in-degree (suggesting the overall usage) information. Hashtags such as \#covid-19, 
\#lockdown, \#stayhome, \#morrison are common to multiple states; therefore, such hashtags are assigned to states where they were prominent. The mentions network also follows the same notion.

\begin{table}[t!]
    \centering
    \caption{Most common hashtags in Australia during the COVID-19 pandemic. In-degree suggests how many states used the hashtag, weighted in-degree suggests the overall volume, and a hashtag assigned to a particular state suggests that the hashtag was used prominently in that state compared to the rest.}
    \begin{tabular}{>{\centering}m{1.5cm}|p{13.5cm}}
    \hline
       \textbf{State}  & \textbf{\#hashtags} with (in-degree:weighted in-degree) information\\
         \hline
       \multirow{3}{*}{VIC}  & \textbf{covid19} (8:29,646), \textbf{lockdown} (8:5,294), \textbf{covid19aus} (8:3,702), \textbf{covid19vic} (8:3,479), \textbf{stayhome} (8:2,842), \textbf{melbourne} (8:2,548), \textbf{stayathome} (8:1,627), \textbf{covidiots} (8:1,411), \textbf{staysafe} (8:1,376), \textbf{covid--19} (8:1,370)\\ \hline
       
       \multirow{3}{*}{NSW} & \textbf{coronavirus} (8:11,033), \textbf{covid} (8:5,348), \textbf{australia} (8:2,996), \textbf{covid19australia} (8:2,378), \textbf{socialdistancing} (8:2,191), \textbf{covid19nsw} (8:1,967), \textbf{pandemic} (8:1,561), \textbf{sydney} (8:1,504), \textbf{quarantine} (8:1,146), \textbf{wfh} (8:1,114) \\\hline
       
       \multirow{2}{*}{QLD} &  \textbf{auspol} (8:7,870), \textbf{workfromhome} (8:590), \textbf{brisbane}(8:391), \textbf{qldpol} (8:346), \textbf{trump} (8:334), \textbf{queensland} (7:325), \textbf{sarscov2} (6:311), \textbf{usa} (8:301), \textbf{covid19qld} (7:285), \textbf{love} (7:284)\\\hline
       
       \multirow{2}{*}{WA} & \textbf{perth} (6:301), \textbf{wapol} (6:220), \textbf{delta} (7:217), \textbf{perthnews} (3:198), \textbf{morrison} (7:196), \textbf{wanews} (3:158), \textbf{covid19wa} (5:125), \textbf{wa} (8:118), \textbf{4corners} (8:105), \textbf{australians} (7:104)\\ \hline
       
       \multirow{2}{*}{SA} &  \textbf{adelaide} (8:391), \textbf{7news} (6:256), \textbf{southaustralia} (5:196), \textbf{covid19sa} (6:136), \textbf{saparli} (4:94), \textbf{covidsa} (4:69), \textbf{sa} (7:65), \textbf{children} (6:62), \textbf{bullshitboy} (4:58), \textbf{salockdown} (2:42)\\\hline
       
       \multirow{2}{*}{ACT} & \textbf{breaking} (8:623), \textbf{canberra} (7:310), \textbf{trumpvirus} (6:170), \textbf{wtf} (8:121), \textbf{qt} (7:111), \textbf{cbr} (2:53), \textbf{canberralockdown} (4:52), \textbf{actlockdown} (3:47), \textbf{thailand} (7:46), \textbf{zerocovid} (6:40) \\\hline
       
       \multirow{2}{*}{TAS} & \textbf{travel} (8:316), \textbf{politas} (6:192), \textbf{covid19tas} (4:156), \textbf{tourism} (8:123), \textbf{tasmania} (6:96), \textbf{hobart} (8:73), \textbf{aviation} (6:57), \textbf{smhr22} (1:32), \textbf{flying} (4:32), \textbf{emergency} (6:29) \\\hline
       
       \multirow{3}{*}{NT} & \textbf{covid19vaccine} (8:149), \textbf{darwin} (6:53), \textbf{coronapocalypse} (7:45), \textbf{closetheschools} (7:43), \textbf{territorytogether} (1:34), \textbf{palmerstonnt} (1:32), \textbf{nt} (5:32), \textbf{shuttheschools} (6:31), \textbf{northernterritory} (6:25), \textbf{thoughtoftheday} (3:23)\\
         \hline
    \end{tabular}
    \label{hashtags-all-states}
\end{table}


\begin{table}[t!]
    \centering
    \caption{Hashtags that were state-specific, i.e. in-degree=1. Notes: $^a$represents total number of hashtags with in-degree=1, and $^b$represents weighted in-degree.}
    \begin{tabular}{>{\centering}m{1.5cm}|p{10.5cm}|p{1cm}|p{1cm}}
    \hline
       \textbf{State}  & \textbf{\#hashtags} & \textbf{\#}$^a$  & \textbf{Wgt.}$^b$\\
         \hline
       VIC  & railphotography, railfansofinstagram, yarratrams, tramsoninstagram & 20,604 & 29,438\\ \hline
       
       NSW & shanghailockdown, mydogposts, oliverscampaign, greatersydneylockdown & 17,567 & 24,677\\\hline
       
       QLD &  surfphotography, surflife, wavesfordays, surfinglife, surfrider & 7,316 & 9,574\\\hline
       
       WA & infreo, perthtwins, rollupforwa, canon5d, keepthebordersclosed & 3,290 & 4,057\\ \hline
       
       SA & adlfest, fiveaa, justbekind, fostercare, kinshipcare & 2,641 & 3,396 \\\hline
       
       ACT & sideastart, lockdownvinyl, cbrlockdown, canberragardener, politicslive & 1,240 & 1,618 \\\hline
       
       TAS &  smhr22, proudtobepublic, covid19pakistan, mona, ecodye & 603 & 734\\\hline
       
       NT & territorytogether, palmerstonnt, sleevesupnt, darwinaustralia & 624 & 808\\
         \hline
    \end{tabular}
    \label{hashtags-single-states}
\end{table}


\begin{table}[t!]
    \centering
    \caption{Most mentions (top 50) in Australia during the COVID-19 pandemic. Notes: $^a$represents the state where the mention was prominent, $^b$represents in-degree of the mention as a node, $^c$represents weighted in-degree.}
    \label{most-mentions}
    \begin{tabular}{>{\centering}m{3.4cm}|c|c|c|>{\centering}m{3.4cm}|c|c|c}
    \hline
    \textbf{@mention} & \textbf{Sig.}$^a$ & \textbf{in-deg.}$^b$ & \textbf{Wgt.}$^c$ & \textbf{@mention} & \textbf{Sig.}$^a$ & \textbf{in-deg}$^b$ & \textbf{Wgt.}$^c$\\
    \hline
scottmorrisonmp & NSW & 8 & 6,997 & 9newsaus & NSW & 8 & 951\\
danielandrewsmp & VIC & 8 & 5,900 & patskarvelas & VIC & 8 & 882\\
gladysb & NSW & 8 & 4,109 & prguy17 & VIC & 8 & 854\\
abcnews & NSW & 8 & 3,605 & australian & VIC & 8 & 839\\
realdonaldtrump & VIC & 8 & 3,470 & who & NSW & 8 & 833\\
albomp & NSW & 8 & 2,151 & lesstonehouse & QLD & 8 & 783\\
nswhealth & NSW & 8 & 1,954 & noplaceforsheep & NSW & 8 & 652\\
skynewsaust & VIC & 8 & 1,916 & 9newsmelb & VIC & 7 & 652\\
covid\_australia & QLD & 8 & 1,892 & mikecarlton01 & NSW & 8 & 640\\
annastaciamp & QLD & 8 & 1,796 & drtedros & NSW & 6 & 639\\
greghuntmp & VIC & 8 & 1,753 & peterfitz & NSW & 8 & 636\\
victoriancho & VIC & 8 & 1,529 & vicgovdhhs & VIC & 6 & 635\\
markmcgowanmp & WA & 8 & 1,347 & abcmelbourne & VIC & 7 & 623\\
theage & VIC & 8 & 1,249 & bradhazzard & NSW & 8 & 608\\
newscomauhq & NSW & 8 & 1,231 & rafepstein & VIC & 8 & 568\\
normanswan & NSW & 8 & 1,182 & afl & VIC & 8 & 557\\
smh & NSW & 8 & 1,152 & abc730 & VIC & 8 & 552\\
vicgovdh & VIC & 7 & 1,150 & domperrottet & NSW & 7 & 550\\
sbsnews & WA & 8 & 1,137 & 7newsmelbourne & VIC & 7 & 535\\
youtube & NSW & 8 & 1,075 & samanthamaiden & NSW & 8 & 530\\
breakfastnews & VIC & 8 & 1,025 & sophieelsworth & VIC & 7 & 530\\
joshfrydenberg & VIC & 7 & 1,018 & timsmithmp & VIC & 8 & 524\\
mjrowland68 & VIC & 8 & 980 & 3aw693 & VIC & 6 & 506\\
theheraldsun & VIC & 8 & 965 & billbowtell & NSW & 8 & 502\\
vanonselenp & NSW & 8 & 955 & leighsales & VIC & 8 & 490\\
\hline
\end{tabular}
\end{table}

Results from hashtags network show a significant presence of state-specific hashtags (i.e., in-degree=1); for instance, Victoria had 70\%, New South Wales had 67\%, and Queensland had 56\% state-specific hashtags. Although the respective weighted in-degrees of the state-specific hashtags are lower, their combined presence in the network is significant. Therefore, information systems for epidemic/pandemic management can benefit by starting with a small set of prominent hashtags such as the ones in Table \ref{keywords-hashtags} and adding state-specific hashtags incrementally for comprehensive data collection and timely identification of region-specific concerns. Similarly, results from the mentions network show that accounts belonging to politicians, government bodies, news channels, journalists, radio stations, social activists, public health officials, and health agencies generate the most engagement. Replies to tweets from and tweets with mentions of such engaging accounts seem to include statements of approval, criticism, and request for aid/volunteering, which after filtration of irrelevant content are advantageous for sketching first-hand reports of a situation as it unfolds. Such accounts can also play a vital role during a pandemic in the dissemination of factual information and diminish the flow of misinformation.

\subsubsection{Tracking of Australian events and their sentiments}

\textbf{Tracking of Australian events.} There were numerous additional topics discussed by the Australian public besides the film industry, economy, finance, workforce, and protests. The timely identification of such topics through social media discussions can be useful in acquiring a better picture of a situation, such that first responders and decision-makers can formulate actionable plans accordingly. In textual data mining, topic modeling is a powerful technique for discovering a set of abstract ``topics'' from a collection of textual documents where each topic represents an interpretable semantic concept. In this study, the abstract ``topics'' are the ``events'' we seek to extract. Topic models can also assist in the screening of tweets specific to humanitarian assistance tasks, such as identifying the demands of a crisis-hit community~---~the discussions related to ``demands'' can be further analyzed for planning the timely distribution of relief supplies. We also perform dynamic topic modeling to analyze the evolution of selected topics over time. Topic modeling in a near-real-time scenario can be effectively used as an event detection technique, and tracking the evolution of a topic helps to understand the trend of word usage related to an event.

For the topic modeling task, we used BERTopic (\cite{grootendorst2022bertopic}) to take advantage of contextual embeddings from sentence transformers (\cite{reimers2019sentence}) and create clusters of topics through a class-based term frequency-inverse document frequency approach (\cite{grootendorst2022bertopic}). We experimented with two settings for \textit{minimum cluster size}: (i) size of $10$ for generating highly-specific clusters, and (ii) size of $1000$ for generating generalized clusters.

\begin{figure}[h!]
    \centering
    \includegraphics[width=0.75\textwidth]{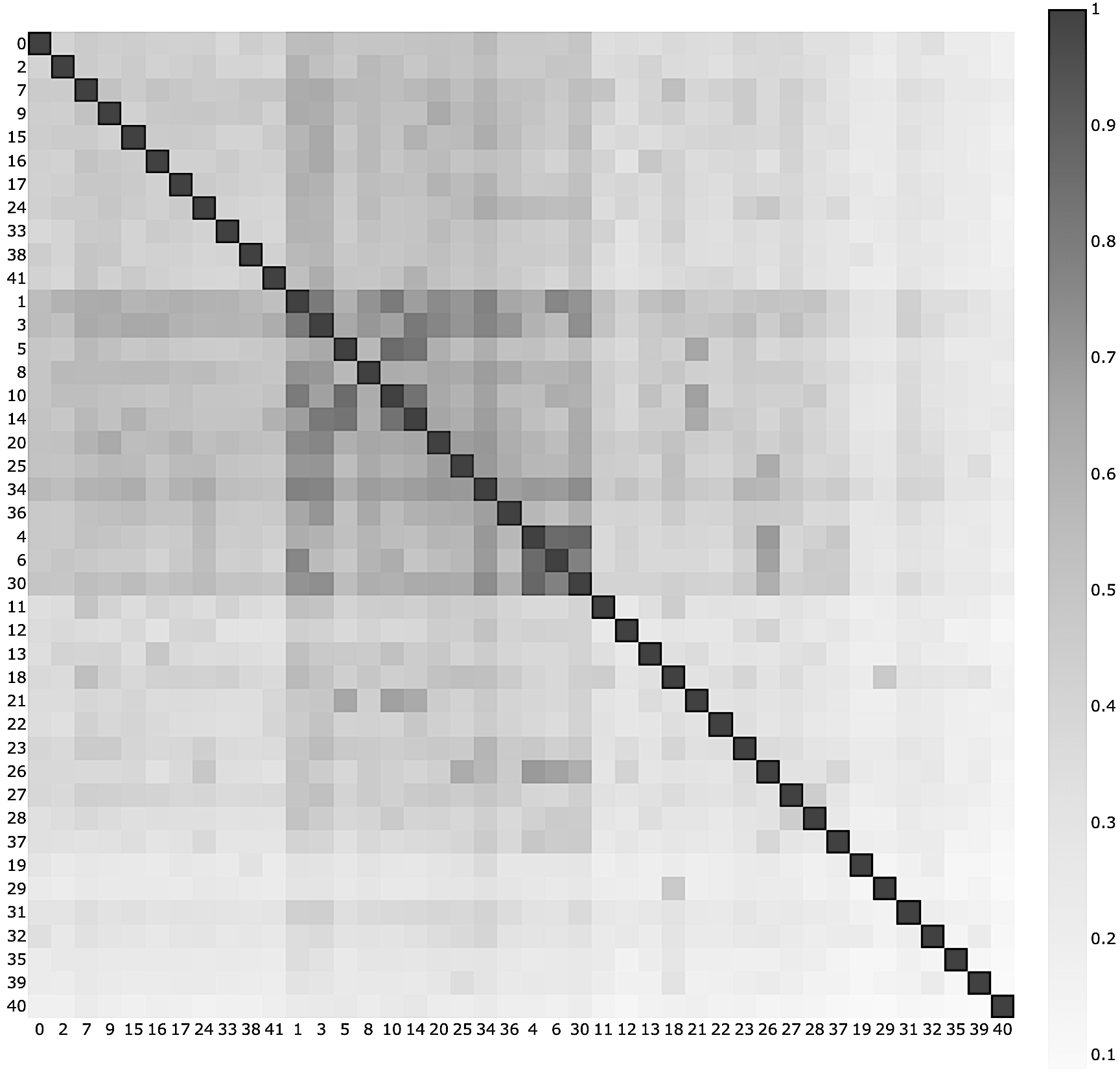}
    \caption{Similarity matrix based on the Cosine similarity. $X$- and $Y$-axis represent topics. Highly correlated topics appear near each other
in the matrix forming dark-colored blocks. Color scale represents similarity score. Refer to Table \ref{aus-general-topics} for topic names.}
\label{similarity-matrix}
\end{figure}

\begin{figure}[t!]
\centering
    \begin{minipage}[t]{.49\textwidth}
        \centering
        \includegraphics[width=\textwidth]{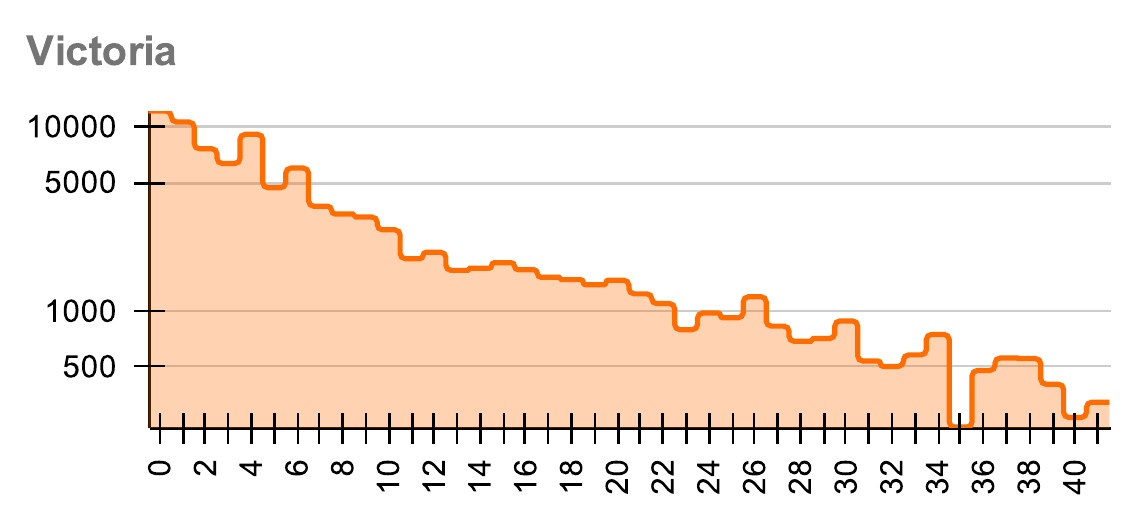}
    \end{minipage}
    \begin{minipage}[t]{.49\textwidth}
        \centering
        \includegraphics[width=\textwidth]{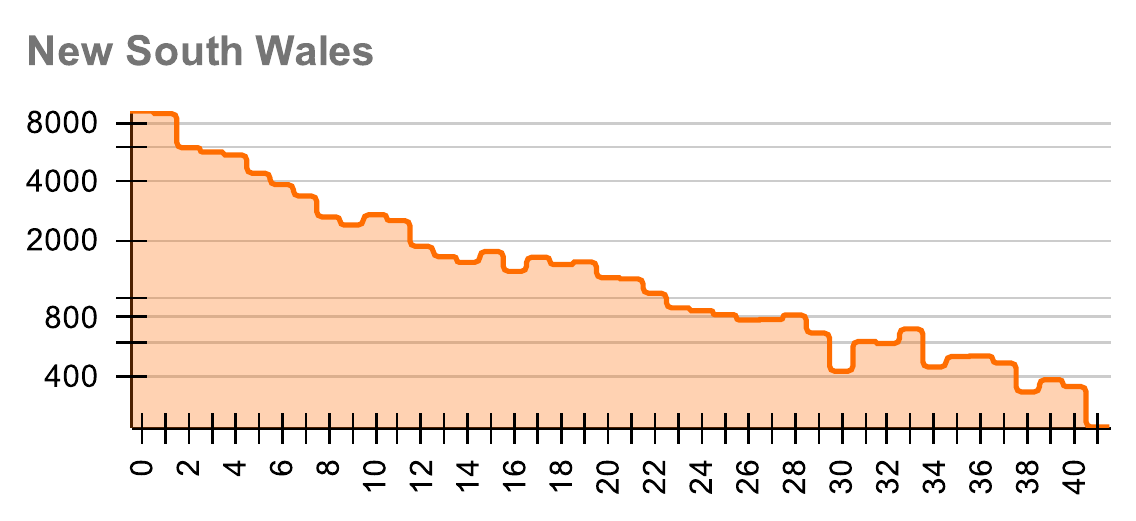}
    \end{minipage}

    \begin{minipage}[t]{.49\textwidth}
        \centering
        \includegraphics[width=\textwidth]{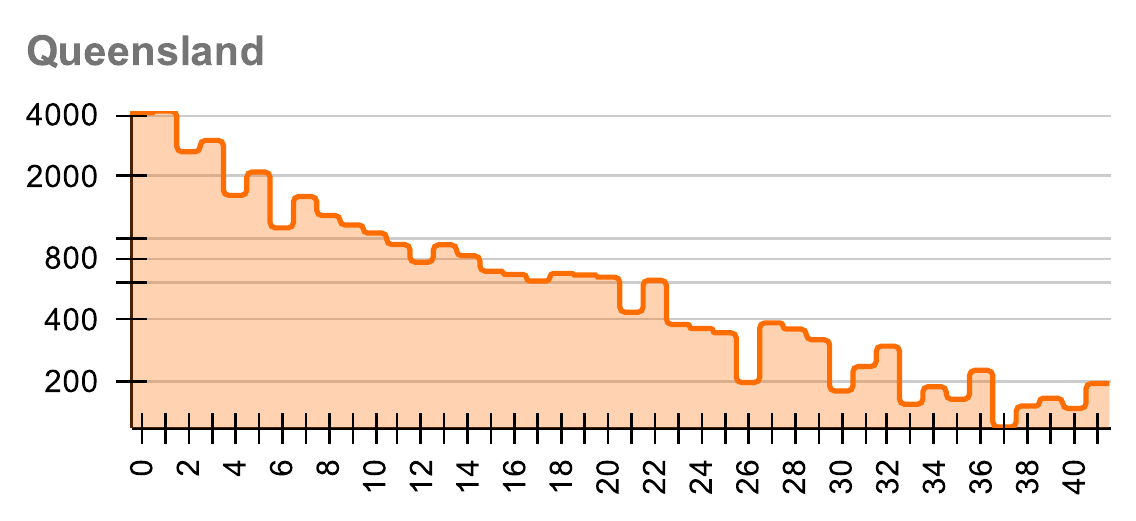}
    \end{minipage}
    \begin{minipage}[t]{.49\textwidth}
        \centering
        \includegraphics[width=\textwidth]{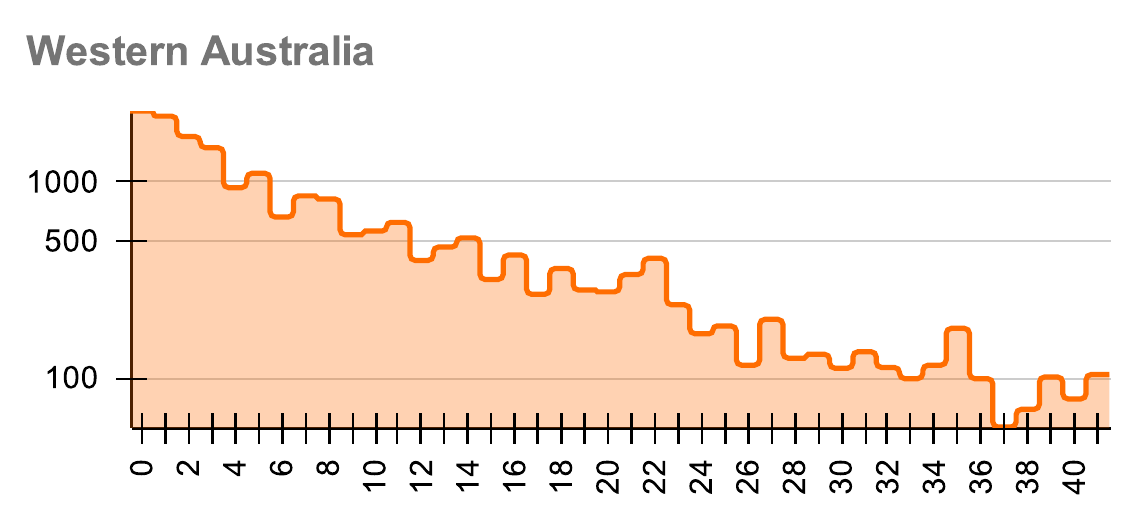}
    \end{minipage}   

    \begin{minipage}[t]{.49\textwidth}
        \centering
        \includegraphics[width=\textwidth]{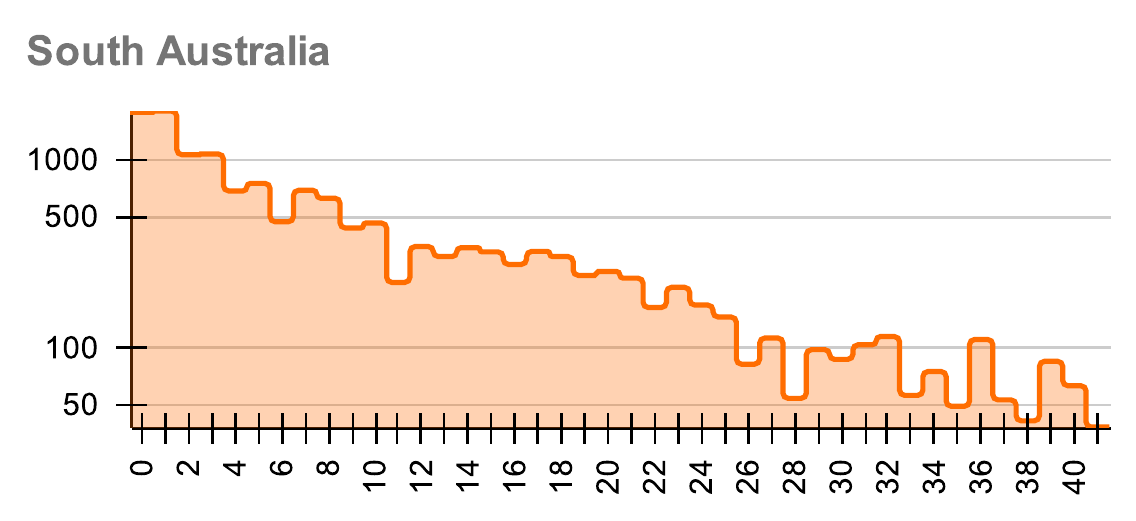}
    \end{minipage}
    \begin{minipage}[t]{.49\textwidth}
        \centering
        \includegraphics[width=\textwidth]{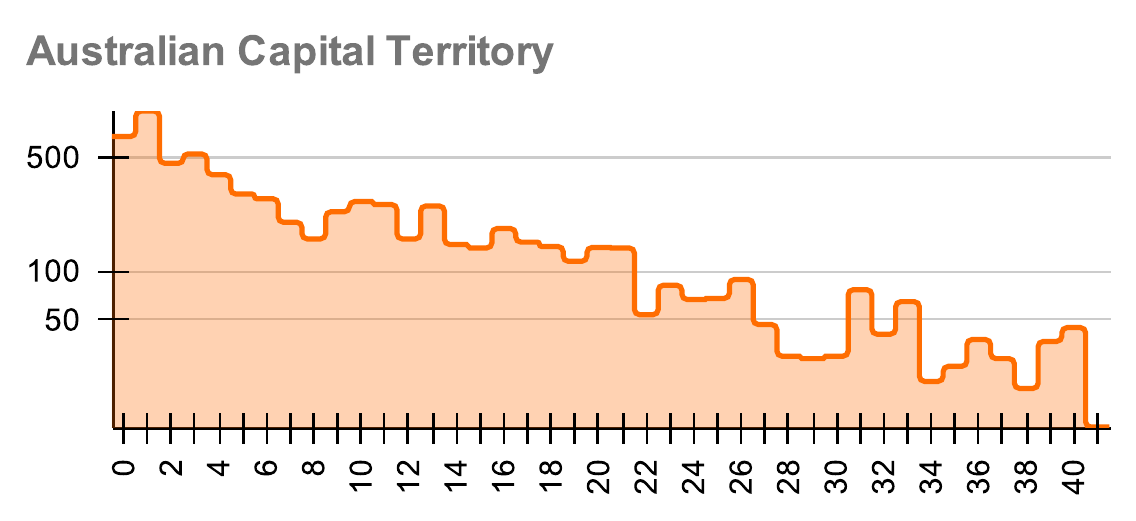}
    \end{minipage}

    \begin{minipage}[t]{.49\textwidth}
        \centering
        \includegraphics[width=\textwidth]{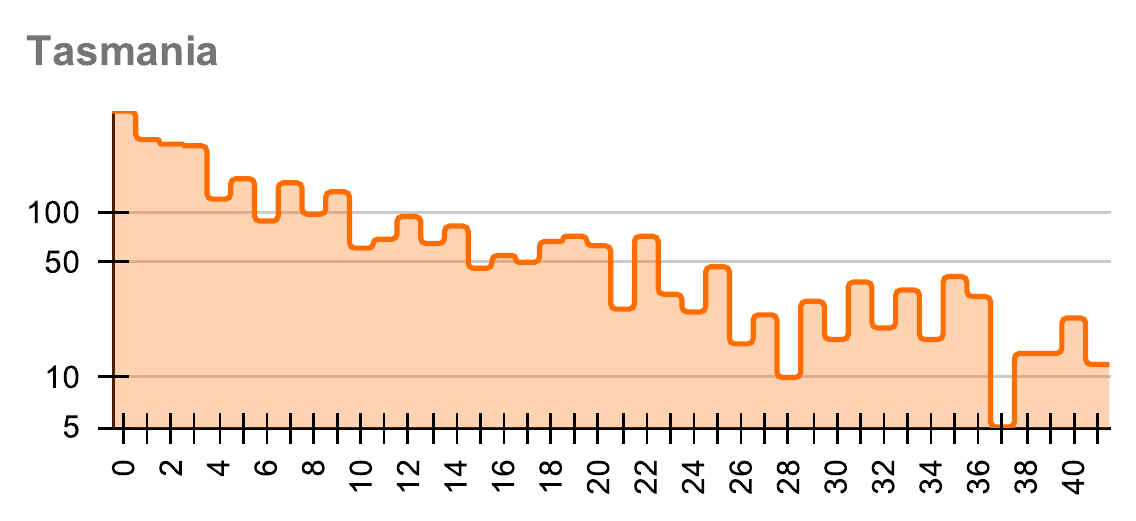}
    \end{minipage}
    \begin{minipage}[t]{.49\textwidth}
        \centering
        \includegraphics[width=\textwidth]{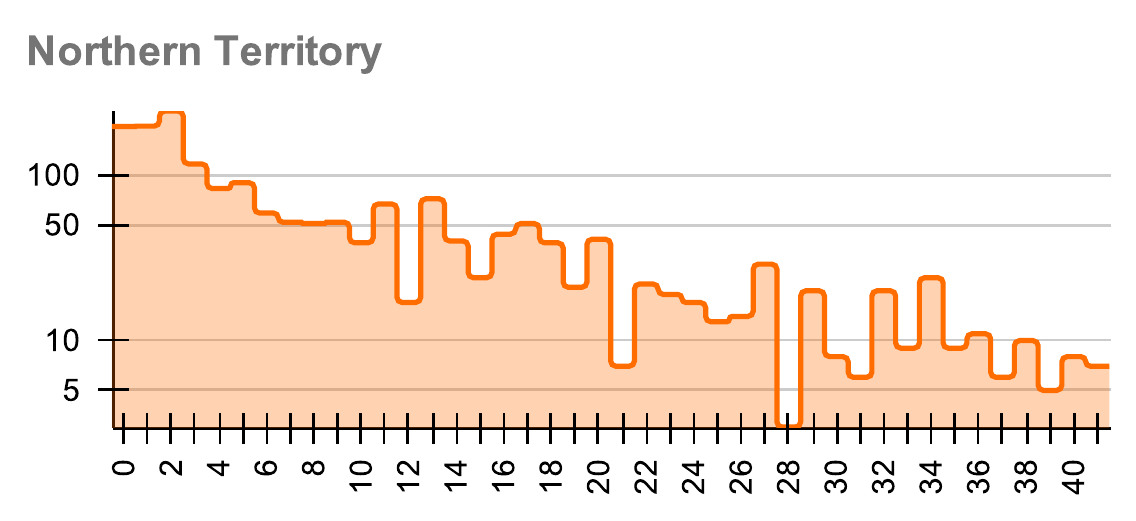}
    \end{minipage} 
     
    \caption{State-topic-based distribution of tweets. For all sub-figures, $Y$-axis represents the number of tweets and is in log scale, and $X$-axis represents topics. For topic names refer to Table \ref{aus-general-topics}.}
    
\label{state-topic-dist}
\end{figure}

Results from highly-specific clusters show that discussions regarding quarantine, isolation, social distancing, illness, Australian politics, the COVIDSafe app, media, vaccines, sports, TV shows and movies, the second wave, restrictions, and shopping, among others, generated the highest interests. More than 2500 topics were identified, out of which topics with at least 450 tweets are listed in Table \ref{highly-specific-clusters}. The listed topics are highly-specific—for instance, there were seven clusters related to discussions on vaccines, namely ``Vaccine rollout", ``Opinions about vaccine", ``Vaccines and vaccinations", ``Vaccines in NSW", ``Vaccine distribution", ``Discussion about the AstraZeneca vaccine", and ``The Pfizer/BioNTech COVID-19 vaccine". Similarly, results from generalized clusters deduced 42 topics, which are listed in Table \ref{aus-general-topics}. Some of the topics in the generalized clusters seem to be highly correlated with each other. We computed cosine similarities of topic embeddings to create a similarity matrix (refer to Figure \ref{similarity-matrix}). While generating the similarity matrix, we cluster the topics such that the highly correlated ones form dense-colored blocks in the matrix. The volumetric patterns in topics for the Australian states and territories in the COVID-19 discourse are identical, with a handful of irregularities. For instance, discussions related to the ``Ruby Princess cruise" were insignificant in Victoria, the issues on ``hairdressing during Lockdown" were insignificant to Queensland, Western Australia, and Tasmania, while discussions around ``Gladys and New South Wales" were negligible in Northern Territory. The state-topic-based tweets distribution is provided in Figure \ref{state-topic-dist}.

\begin{table}[h!]
    \centering
    \caption{Highly-specific clusters-based most discussed topics in Australia during the COVID-19 pandemic. Topics are sorted based on the volume of tweets, and we list topics with at least 450 tweets.}
    \label{highly-specific-clusters}
    \begin{tabular}{p{6cm}|p{6cm}}
    \hline
\textbf{Topic Name} $\downarrow$(0--33) & \textbf{Topic Name} $\downarrow$(34--67) \\ \hline
(Self) Quarantine & Vaccines in NSW  \\
Global pandemic & Pets and other animals during the pandemic\\
The Virus & PPE\\
(Seasonal) Flu & Productivity, labor force, unemployment \\
COVID-19 general discussions &Deaths due to COVID-19 \\
Australian politics & Vaccine distribution\\
Schools and education & Chinese Virus \\
Social distancing & COVID-19 in NSW\\
Prime Minister Scott Morrison & Opinions on COVID-19\\
COVID-19 and live music &  Self-isolation\\
COVID-19 testing & Unproven treatment for COVID-19 \\
COVID-19 and covidsafe app & Mental health issues\\
Lockdown & Masks \\
Work from home & Opinions about Australian Open 2021\\
Lockdown in Melbourne & Coughing and other COVID-19 symptoms\\
Protests & Italian people\\
Vaccine rollout & Discussion about the Astrazeneca vaccine\\
Wearing masks & Coronavirus in India\\
Hotel quarantine & Tenant rights\\
Opinions about vaccines & Rat kits and test performance\\
Sporting events & Caroline Flack\\
Lockdown in Sydney & Universities and COVID-19\\
Astrazeneca and Pfizer & TV and film recommendations\\
Gladys & Lockdown cooking\\
China & About Aged care facilities and COVID-19\\
TV shows and movies &  The Pfizer/BioNTech COVID-19 vaccine\\
COVID-19 variants & Health workers and frontline workers \\
Victorians & World War I and the Spanish flu\\
Shopping and availability of products & Cruise ships and COVID-19\\
Hashtags related to Scott Morrison  & COVID-19 and children\\
Got COVID-19 & Dan Andrews and Coronavirus\\
Flu shots and immunizations & State by state latest updates\\
Vaccines and vaccinations &Easter\\
Ruby Princess & Ashes, Cricket\\
\hline
\end{tabular}
\end{table}

\begin{table}[t!]
    \centering
    \caption{Generalized clusters-based most discussed topics in Australia during the COVID-19 pandemic. Topics are sorted based on the volume of tweets.}
    \begin{tabular}{p{6cm}|p{6cm}}
    \hline
\textbf{Topic Name} & \textbf{Topic Name} \\ \hline
(0) Face mask                   & (21) Astrazeneca and Pfizer                \\ 
(1) NSW and Victoria            & (22) Flu                                   \\ 
(2) Hotel Quarantine                 & (23) Self-promotion, links, spam           \\ 
(3) COVID-19 general discussions     & (24) Music, Albums                    \\ 
(4) Lockdown                         & (25)  Food, shopping                        \\ 
(5) Vaccines                         & (26)  Dinner and cooking in lockdown \\ 
(6) Lockdown in Melbourne and Sydney & (27)  Referring to women figures            \\ 
(7) The Virus                        & (28) Gladys and New South Wales            \\ 
(8) Sporting events             & (29)  Corona                                \\ 
(9) COVID-19 as a pandemic           & (30)  Lockdown in Victoria                  \\ 
(10)  Vaccines in Australian context   & (31)  COVIDsafe app and tracing             \\ 
(11)  China and Wuhan                  & (32)  Hygiene                               \\ 
(12)  Work from home                   & (33)  COVID-19 and India                    \\ 
(13)  Scott Morrison                   & (34)  Wear a mask                           \\ 
(14)  Vaccinated                       & (35)  Cruise ships, infection, and outbreak \\ 
(15)  COVID-19 testing                 & (36)  TV shows and movies                  \\ 
(16)  COVID-19 and Trump               & (37)  Haircuts, hairdressing in Lockdown    \\ 
(17)  Schools and education            & (38)  Protests                              \\ 
(18)  Coronavirus                      & (39)  Toilet paper and panic buying         \\ 
(19)  Social distancing                & (40)  Herd Immunity                         \\ 
(20)  Jobs              & (41) Vax, vaxxed and anti                  \\ \hline
\end{tabular}
\label{aus-general-topics}

\end{table}

We also studied the evolution of keywords through dynamic topic modeling. The study timeline was split into 33 sub-timelines, each representing months between January 1, 2020, and October 9, 2022. We summarize the results from the analysis for selected topics (due to space limitations), namely ``Face masks", ``Lockdown in Australia", ``Sporting events", ``Vaccines in Australian context", ``Jobs", and ``Toilet paper and panic buying", in Table \ref{aus-dynamic-model}. The listed sets of keywords are influential terms describing the respective topics during the mentioned timeline.

\begin{table}
    \centering
    \caption{Evolution of keywords for selected topics. We list only selected timelines due to space limitations. Keywords are
influential terms describing the respective topics during the mentioned timeline.}
    \begin{tabular}{p{2.5cm}|p{11.5cm}}
    \hline
      \textbf{Topic Name}  & \textbf{Keywords} \\ \hline
\multirow{2}{2cm}{\textbf{Hotel \newline Quarantine}} &
\textbf{January 2020}: quarantine, evacuees, 14, days
\textbf{March 2020}: quarantine, self
\textbf{June 2020}: quarantine, hotel
\textbf{April 2022}: quarantine, travellers, inbound, redundant     \\ \hline

\multirow{6}{2cm}{\textbf{Lockdown in Australia}} & 
\textbf{March 2020}: stayhome, australia, lockdown, sydney, melbourne
\textbf{July 2020}: melbourne, lockdown, victoria, stage
\textbf{November 2020}: lockdown, melbourne, adelaide, australia, south
\textbf{Janurary 2021}: lockdown, brisbane, perth, 6pm
\textbf{March 2021}: brisbane, lockdown, queensland
\textbf{April 2021}: perth, lockdown, perthlockdown
\textbf{November 2021}: lockdown, melbourne, back
\textbf{April 2022}: lockdown, melbourne, longest
\textbf{August 2022}: rent, lockdown, nsw, longest, vic
\\ \hline

\multirow{4}{2cm}{\textbf{Sporting events}} &
\textbf{January 2020}: coronavirus, olympics, tokyo
\textbf{March 2020}: coronavirus, afl, nrl, season
\textbf{June 2020}: afl, essendon, game
\textbf{December 2020}: cricket, scg, test, players
\textbf{January 2021}: tennis, players, quarantine, cricket
\textbf{July 2021}: olympics, nrl, players, afl
\textbf{December 2021}: ashes, djokovic, novak, tennis, covid
\\ \hline

\multirow{5}{2cm}{\textbf{Vaccines in Australian context}} &
\textbf{March 2020}: vaccine, australia, testing
\textbf{April 2020}: vaccine, australia, until, vahs
\textbf{January 2021}: vaccine, australia, vaccines, pfizer
\textbf{February 2021}: vaccine, australia, vaccines, rollout
\textbf{August 2021}: nsw, vaccine, vaccines, vaccinated, vaccination
\textbf{December 2021}: vaccine, booster, vaccinated, vaccines, australia
\textbf{March 2022}: fourth, vaccine, australia, rolled, updates
\\ \hline

\multirow{6}{2cm}{\textbf{Jobs}} &
\textbf{January 2020}: coronavirus, recession, global, leads
\textbf{March 2020}: workers, coronavirus, health
\textbf{May 2020}: economy, workers, nurses
\textbf{June 2020}: economy, unemployment, recession, jobs
\textbf{December 2020}: bill, covid, workers, relief
\textbf{January 2021}: covid, jobkeeper, money, workers
\textbf{April 2021}: jobseeker, keeper, supplement, covid, payments
\textbf{July 2021}: covid, workers, health, pay
\textbf{January 2022}: workers, economy, health, covid, supply
\textbf{April 2022}: covid, unemployment, labor, pandemic
\\ \hline

\multirow{2}{2cm}{\textbf{Panic buying}} & \textbf{January 2020}: paper, toilet, tootpapercrisis2020
\textbf{March 2020}: toilet, paper, coronavirus, toiletpaper, buying
\textbf{June 2020}: toilet, paper, hoarding, panic, buying
\\ \hline

\end{tabular}
\label{aus-dynamic-model}

\end{table}

\textbf{Learning the sentiments of people.} People share their opinions and feelings regarding various dynamics of a crisis event. The outbreak followed by lockdowns, curfews, travel restrictions, social distancing, quarantine, and a cumulative rise in confirmed cases and deaths between 2020--2022 affected people both in terms of physical and mental health. Studies related to the pandemic and sentiments have reported a rise in negative feelings and pessimism. Therefore, we investigate the overall sentiment trend of the Australian Twitterverse during different phases of the pandemic. Australia experienced four major COVID-19 waves\footnote{https://www.abs.gov.au/articles/covid-19-mortality-wave}: (i) March--May 2020, (ii) June--November 2020, (iii) July--December 2021 (Delta wave), and (iv) during 2022--until the end of September 2022 (Omicron wave).

\begin{figure}
    \centering
    \includegraphics[width=0.98\textwidth]{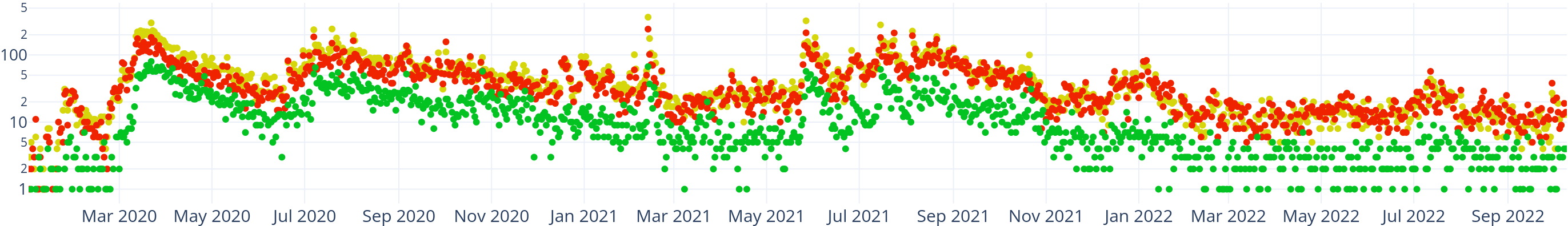}
    \caption*{(a) VIC}
    \includegraphics[width=0.98\textwidth]{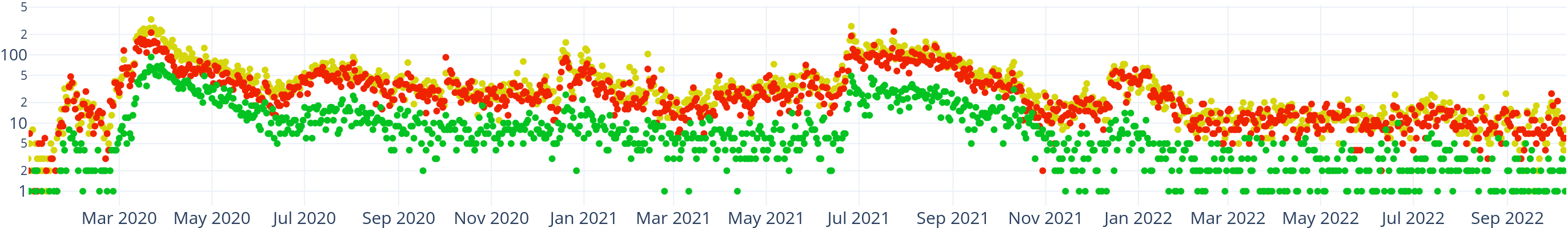}
    \caption*{(b) NSW}
    \includegraphics[width=0.98\textwidth]{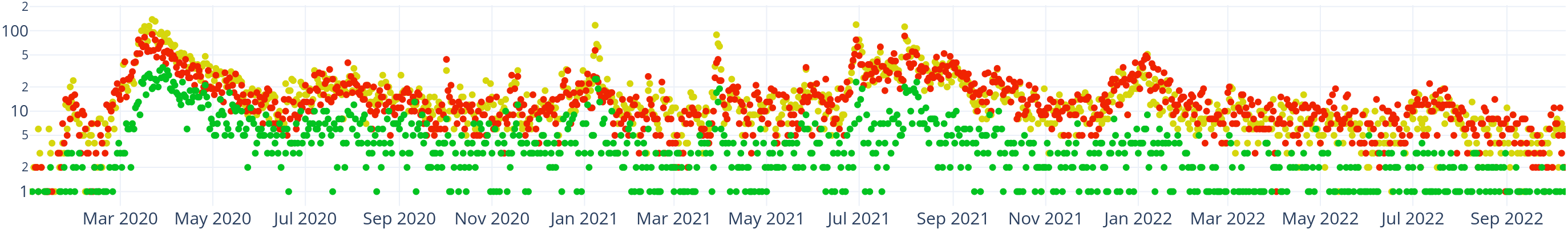}
    \caption*{(c) QLD}
    \includegraphics[width=0.98\textwidth]{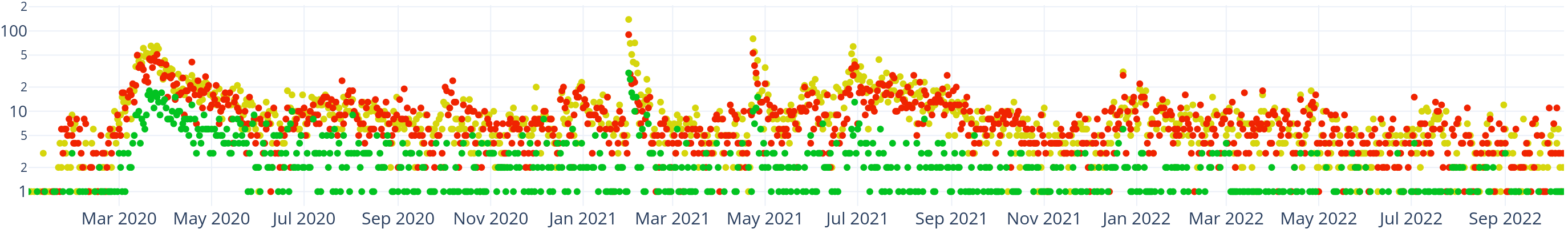}
    \caption*{(d) WA}
    \includegraphics[width=0.98\textwidth]{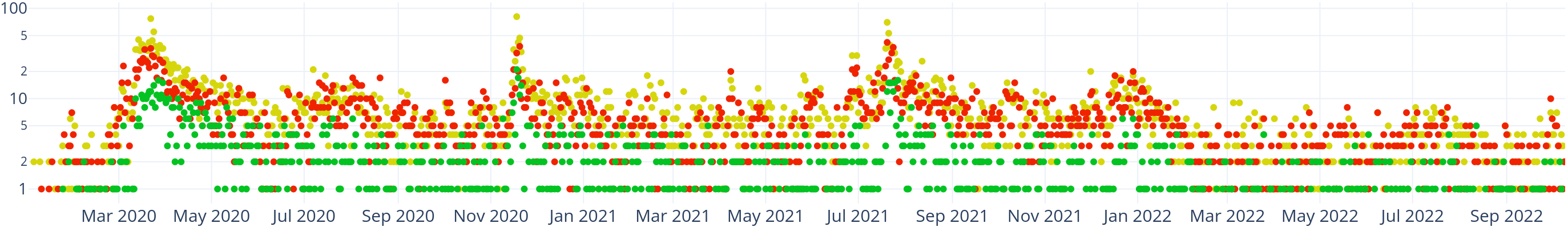}
    \caption*{(e) SA}
    \includegraphics[width=0.98\textwidth]{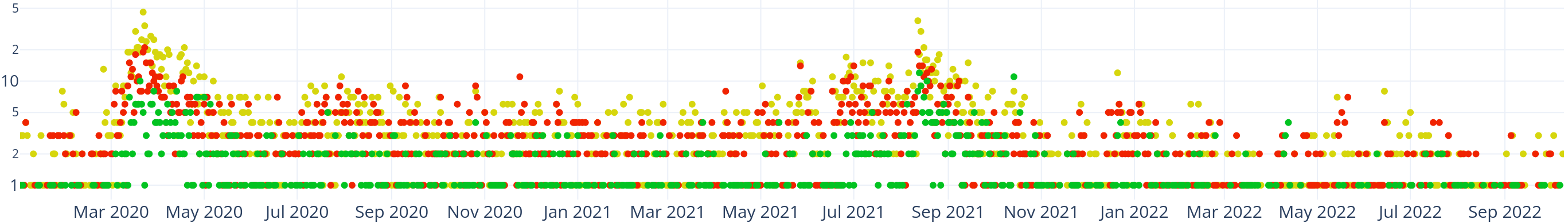}
    \caption*{(f) ACT}
    \includegraphics[width=0.98\textwidth]{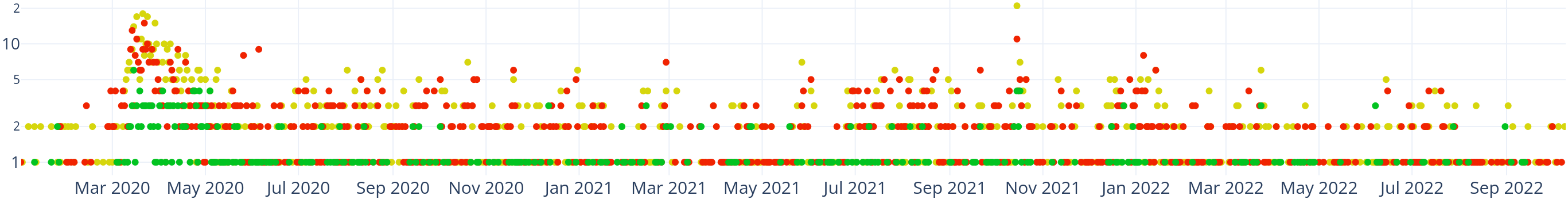}
    \caption*{(g) TAS}
    \includegraphics[width=0.98\textwidth]{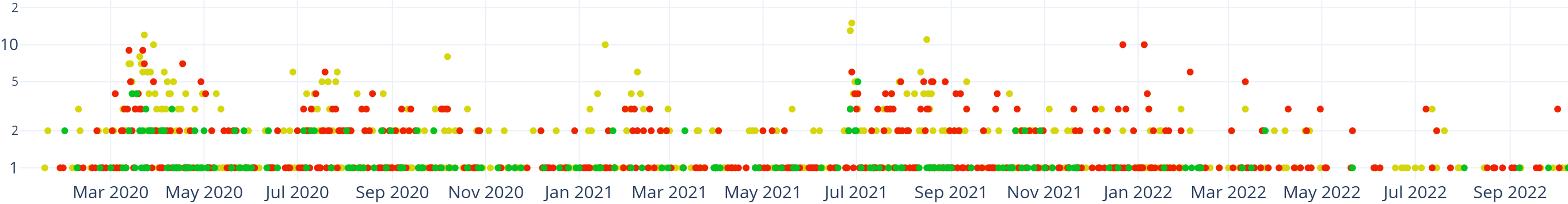}
    \caption*{(h) NT}
    \caption{Daily distribution of tweets based on sentiments. $Y$-axis represents the number of tweets and is in log scale. Yellow, Red, and Green dots represent neutral, negative, and positive sentiments, respectively.}
    \label{daily-sentiments-states}
\end{figure}

\begin{landscape}
\begin{figure}
    \centering
    \includegraphics[width=23cm]{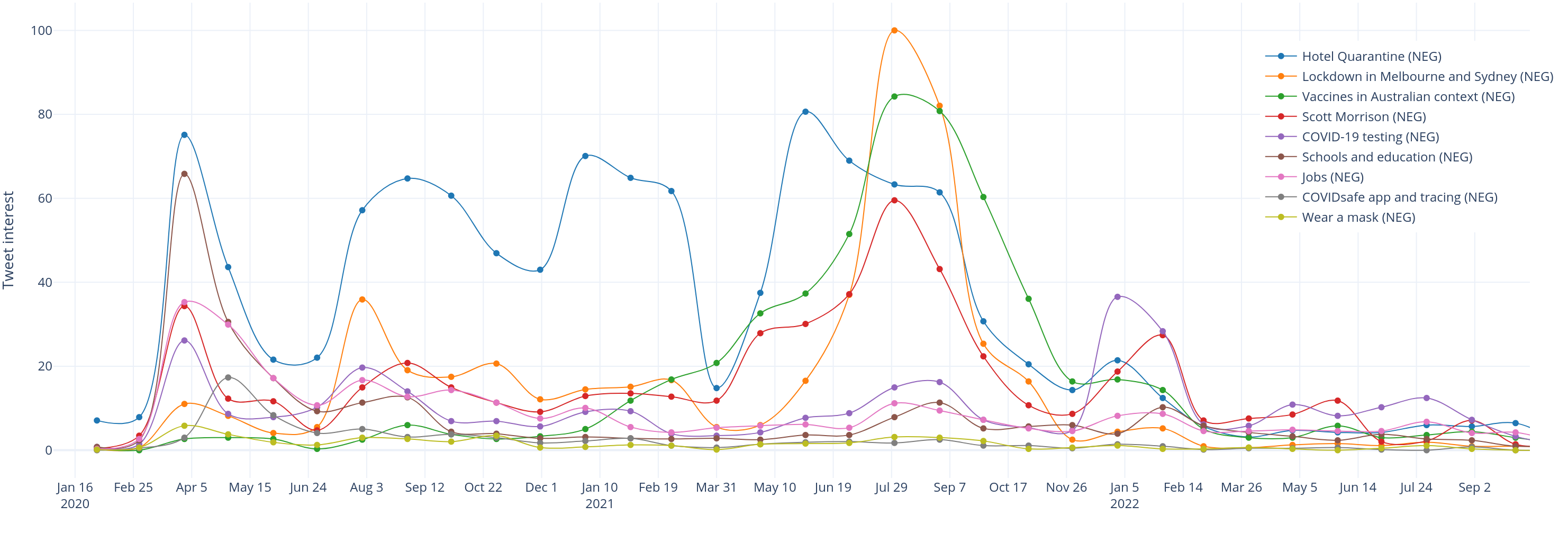}
    \includegraphics[width=23cm]{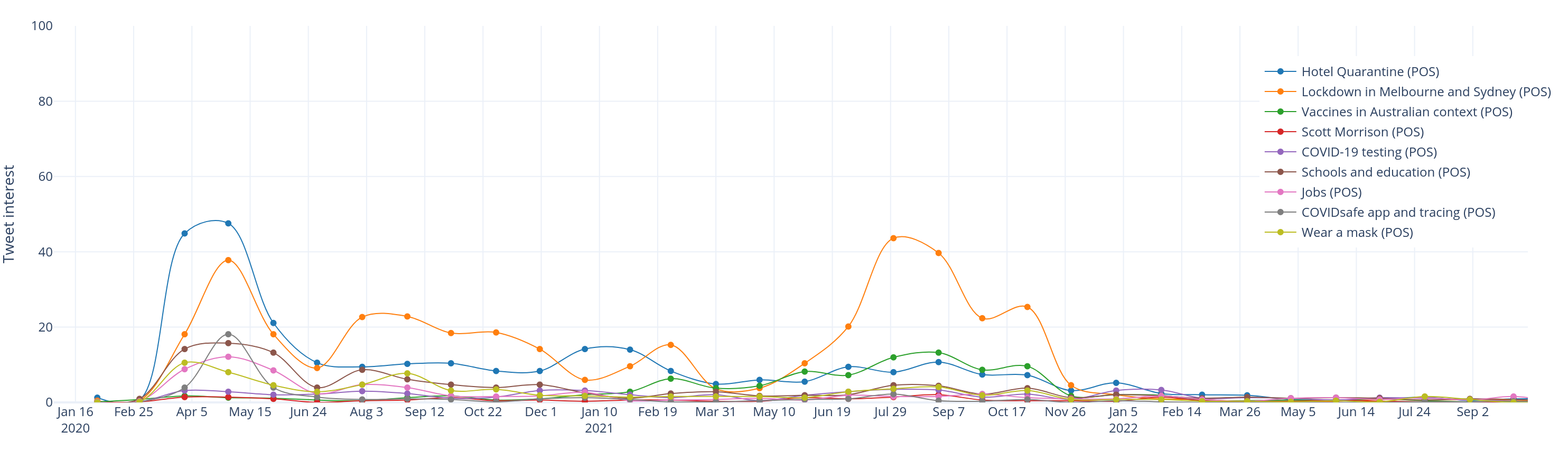}
    \caption{Sentiment analysis on selected topics. The top plot refers to the distribution of negative sentiment tweets, and the bottom plot refers to the distribution of positive sentiment tweets. $Y$-axis is tweet interest~---~numbers represent interest relative to the highest point on the first plot; therefore, the two plots are comparable.}
    \label{topic-wise-sentiments}
\end{figure}
\end{landscape}

For the sentiment analysis task, we finetuned BERTweet (covid19-base-cased) on SemEval-2017 Task 4 dataset (\cite{rosenthal-etal-2017-semeval}). The results from sentiment analysis are summarized in Figure \ref{daily-sentiments-states} and Figure \ref{topic-wise-sentiments}. The neutral, negative, and positive sentiment brackets for each state and territory were as follows: Australian Capital Territory [50.98\%, 34.80\%, 14.22\%], New South Wales [49.36\%, 39.01\%, 11.63\%], Northern Territory [48.36\%, 39.95\%, 11.69\%], Queensland [46.29\%, 42.25\%, 11.46\%], South Australia [49.12\%, 38.61\%, 12.27\%], Tasmania [46.91\%, 41.19\%, 11.90\%], Victoria [47.45\%, 39.82\%, 12.73\%], Western Australia [47.71\%, 42.20\%, 10.10\%]. The daily distribution of tweets across states based on their sentiments is shown in Figure \ref{daily-sentiments-states}. There is the presence of significant peaks in tweet interest across all states during the first three waves. The interest, however, does not seem associated with the fourth wave, except for Victoria, where neutral and negative tweets form a peak for a restricted timeline. Overall, the Australian Twitterverse seemed more inclined toward neutral and negative sentiments.

To have a better perspective on the pandemic sentiments, we selected a few topics (due to space limitations) and explored their interests over time. Figure \ref{topic-wise-sentiments} presents sentiment analysis on the selected topics, namely ``Hotel Quarantine", ``Lockdown in Melbourne and Sydney", ``Vaccines in Australian context", ``Scott Morrison", ``COVID-19 testing", ``Schools and education", ``Jobs", ``COVIDsafe app and tracing", and ``Wear a mask". Tweet interests in Figure \ref{topic-wise-sentiments} are computed as relative to the highest point in the plots, similar to the search trends analogy of Google trends. Results show that negative sentiments majorly dominated the discourse. Lockdown-related discussions had more positive sentiments during the first wave; as negative sentiments started to become significant during the early month of the second wave, the topic recorded the highest tweet interest during the third wave. Discussions on hotel quarantine had negative sentiments throughout 2020--early 2022, with statistically significant positive sentiments during the first wave. Discussions on vaccines started gaining tweet interest in early 2021 and attained the second most tweet interest (with negative sentiments) during the third wave. Similarly, discussions related to Scott Morrison with negative sentiments achieved significantly high tweet interest during the third wave. Tweets related to COVID-19 testing also inclined significantly towards negative sentiments recording its highest tweet interest in early 2022. After February 2022, the discourse around the selected subjects shows descending tweet interests.

Our study suggests that the number of topics during a pandemic can be in the thousands as country-, state-, city-, county-, and district-level large-scale and small-scale events accumulate over time. Although the topics from highly-specific clustering were based on a minimum topic size of 10, the tweet corpus had only geotagged tweets, and today $<$1\% of tweets are geotagged. A comparative analysis done by (\cite{lamsal2022twitter}) reported the daily distributions of full-volume (based on Twitter's \textit{counts API}) and geotagged tweets to have significantly identical patterns; therefore, the identified topics are near-true representations of the events discussed during the pandemic in Australia. Hence, a minimum topic size of 10 helps identify small-scale events. The identification of small-scale events is necessary as they include the concerns of a district or a county. Overall, topical analysis shows that information systems can largely assist management authorities in obtaining a comprehensive situational view of an epidemic or pandemic through the mining of highly-specific topics and performing further analyses~---~tracking evolutions of keywords, exploring sentiment trends, and studying tweeting interests and causality behavior~---~inside the topic clusters.

\subsubsection{Studying the causality behavior of discussion-based time series}
Geotagged Twitter discussions have been reported to have latent variables (time series) that Granger-cause the daily confirmed COVID-19 cases (\cite{lamsal2022twitter}). With a causality analysis of discussion-based time series, we seek to identify a set of time series that Granger-cause the Australian confirmed cases and death cases. For this task, we generated multiple time series based on the volume of tweets over different topics and sentiments. The generated time series dataset took the following form:

\begin{equation*}
\label{tsdataset}
Time~series: T_{tp^{j}sn^{k}}^{t^{i}}
\end{equation*}
Where, $t^i$ represents the date component, $tp^j$ represents the topic component, and $sn^k$ represents the sentiment component.

Consider a time series $y$. Its autoregressive model $y_{t}$ is:

\begin{equation}
\label{autoreg}
    y_{t} =  a_{0}+a_{1}y_{t-1}+a_{2}y_{t-2}+...+a_{n}y_{t-n}+ e_{t}
\end{equation}

Consider another time series $x$. Now we include the lagged values of $x$ into Equation \ref{autoreg}:

\begin{equation}
    \label{autoregfull}
    y_{t} =  a_{0}+a_{1}y_{t-1}+a_{2}y_{t-2}+...+a_{n}y_{t-n}+ b_{s}x_{t-s}+...+b_{l}x_{t-l}+e_{t}
\end{equation}

According to Granger-causality (\cite{Granger}), $x$ Granger-causes $y$ if lagged values of $x$ in Equation \ref{autoregfull} are significant (in \textit{F}-test). We performed causality tests between $y$, i.e., confirmed cases and death cases, and each $tp^{j}sn^{k}$ for the maximum lags of 90 at a 5\% significance level. To identify the variables that Granger-cause the confirmed cases, we considered every $tp^{j}sn^{k}$ as independent variables, while for the death cases we also included the daily confirmed cases as an independent variable besides $tp^{j}sn^{k}$. The results from causality tests are summarized in Table \ref{lag-table-confirmed-cases} and Table \ref{lag-table-death-cases}.

\begin{table}[t!]
    \centering
    \caption{Time series that Granger-cause (at 5\% significance) the Australian confirmed COVID-19 cases.}
    \label{lag-table-confirmed-cases}
    \begin{tabular}{ >{\centering}m{2cm} | >{\centering}m{5cm} |c|c}
    \hline
    \textbf{Time series} & \textbf{Discussion cluster} & \textbf{Signif. $p$-values} & \textbf{Signif. lags}\\
    \hline
    $tp^{3}sn^{NEG}$ & COVID-19 general discussions & 30  & 4--33 \\
    $tp^{8}sn^{NEG}$ & Sporting events & 89 & 1, 3--90 \\
    $tp^{13}sn^{NEG}$ & Scott Morrison & 34 & 2, 9--38, 40--42\\
    $tp^{14}sn^{NEG}$ & Vaccinated & 67 & 18, 25--90\\
    $tp^{15}sn^{NEG}$ & COVID-19 testing & 84 & 7--90\\
    $tp^{25}sn^{NEG}$ & Food, shopping & 3 & 9--11\\
    $tp^{36}sn^{NEG}$ & TV shows and movies & 82 & 1--3, 5, 13--90\\
    $tp^{41}sn^{NEG}$ & Vax, vaxxed and anti & 75 & 8, 16--18, 20--90\\
    $tp^{3}sn^{NEU}$ & COVID-19 general discussions & 27 & 8--29, 40--43, 46\\
    $tp^{8}sn^{NEU}$ & Sporting events & 85 & 6--90\\
    $tp^{14}sn^{NEU}$ & Vaccinated & 8 & 8--13, 15, 29\\
    $tp^{15}sn^{NEU}$ & COVID-19 testing & 88 & 3--90\\
    $tp^{27}sn^{NEU}$ & Referring to women figures & 15 & 9--16, 28, 30--35\\
    $tp^{36}sn^{NEU}$ & TV shows and movies & 76 & 15--90\\
    $tp^{41}sn^{NEU}$ & Vax, vaxxed and anti & 72 & 19--90\\
    $tp^{15}sn^{POS}$ & COVID-19 testing & 49 & 5--53\\
    $tp^{36}sn^{POS}$ & TV shows and movies & 76 & 15--90\\
    $tp^{41}sn^{POS}$ & Vax, vaxxed and anti & 86 & 5--90\\
    \hline
    \end{tabular}
\end{table}

\begin{table}[t!]
    \centering
    \caption{Time series that Granger-cause (at 5\% significance) the Australian COVID-19 death cases.}
    \label{lag-table-death-cases}
    \begin{tabular}{ >{\centering}m{2cm} | >{\centering}m{2.9cm} |c|c}
    \hline
    \textbf{Time series} & \textbf{Discussion cluster} & \textbf{Signif. $p$-values} & \textbf{Signif. lags}\\
    \hline
    $tp^{14}sn^{NEG}$ & Vaccinated & 1  & 33 \\
    $tp^{15}sn^{NEG}$ & COVID-19 testing & 55 & 36--90 \\
    $tp^{41}sn^{NEG}$ & Vax, vaxxed and anti & 2 & 89,90\\
    $tp^{15}sn^{NEU}$ & COVID-19 testing & 17 & 44--46, 48--50, 52, 53, 78, 81, 82, 85--90\\
    $tp^{22}sn^{NEU}$ & Flu & 1 & 9\\
    $tp^{41}sn^{NEU}$ & Vax, vaxxed and anti & 2 & 88,89\\
    $tp^{13}sn^{POS}$ & Scott Morrison & 2 & 72, 73\\
    $tp^{41}sn^{POS}$ & Vax, vaxxed and anti & 48 & 29--56, 62--66, 76--90\\
    $cases$ & - & 90 & 1--90\\
    \hline
    \end{tabular}
\end{table}

Results show the presence of 18 variables (out of 126) that Granger-cause the daily confirmed COVID-19 cases (refer to Table \ref{lag-table-confirmed-cases}). Variables related to Sporting events [negative sentiments], COVID-19 testing [neutral sentiments], Vax, vaxxed and anti [positive sentiments], Sporting events [neutral sentiments], COVID-19 testing [negative sentiments], TV shows and movies [negative sentiments] were observed Granger-causing the confirmed cases for more than 80 lags. Similarly, we identified 9 variables that Granger-cause the Australian COVID-19 death cases (refer to Table \ref{lag-table-death-cases}). Results show that the daily confirmed COVID-19 cases Granger-cause the death cases for all 90 lags, and is followed by COVID-19 testing [negative sentiments] with 55 significant lags, Vax, vaxxed and anti [positive sentiments] with 48 significant lags, COVID-19 testing [neutral sentiments] with 17 significant lags. The variables that Granger-cause the confirmed cases and death cases reveal additional information about their forecasting properties. Most of the variables that Granger-cause the confirmed cases start to provide forecasting power within their 1--2 weeks lag. However, for the variables that Granger-cause the death cases, except for flu and confirmed cases, the explanatory power is evident only after a couple of weeks.

The causality analysis in our study used ``volumetric" features of topics discussed during a pandemic. Using ``volume" as a feature reduces the computational complexity since we rely only on the volume of tweets based on their topical and sentimental characteristics. Inclusion of the Granger-causing variables, such as the ones listed in Table \ref{lag-table-confirmed-cases} and Table \ref{lag-table-death-cases}, into forecasting models, have shown improved performance on forecasts compared to models fitted on just the lagged values of the dependent variable (\cite{lamsal2022twitter}). Pandemic (confirmed and deaths) cases forecasting models fitted on such time series data can be deployed on small-scale infrastructures. Early predictions of the cases help authorities and decision-making bodies to make early estimates of resources to cope with the consequences of future waves of an ongoing epidemic or pandemic.

\section{Conclusion}
During an ongoing crisis, people use social media as a broadcast platform for disseminating situational updates through exchanges of statuses, stories, and media items regarding what they have seen, felt, or heard. Such conversations, if timely monitored and analyzed, can contain actionable information that can assist first responders and decision-makers in formulating plans for effective disaster management. In this study, we performed an extensive analysis of COVID-19-related Twitter discussions generated in Australia between January 2020, and October 2022, and discussed the significance of such analysis towards the extraction of ``situational awareness" concerning a crisis event. We analyzed hashtags and mentions at the state level with in-depth network analysis and performed topic modeling to discover highly-specific topics and generalized topics discussed by the Australian Twitterverse during the pandemic. Next, we explored the conversation dynamics of the Twitterverse across topics and sentiments over temporal and spatial dimensions. Finally, we utilized the knowledge gathered during topic modeling and sentiment analysis to generate numerous discussion-based time series to study the causality behavior of each time series on the Australian COVID-19 confirmed cases and death cases. Overall, we studied the discussion dynamics of the COVID-19 pandemic in Australia to also explore areas that can aid in designing future automated information systems for effective epidemic/pandemic management.

\section*{Acknowledgements}
This study was supported by the \textit{Melbourne Research Scholarship} from the University of Melbourne, Australia. We (the authors) are thankful to \textit{Nectar Research Cloud} (a service of the \textit{Australian Research Data Commons}) for supporting this study with a large-volume compute instance (24 VCPUs, 216GB memory, 20TB volume).

\section{Data Availability}
The data collected as a part of this study is available at \texttt{https://dx.doi.org/10.21227/42h1-ge40} as an open-access item. We name the dataset \texttt{MegaGeoCOV Extended}. A free \textit{IEEE} account is sufficient to access the dataset. The shared tweet identifiers need to be hydrated to re-create the dataset locally. Note that, after hydration, the number of tweets can vary as deleted or private tweets are not retrievable. The dataset includes the following tweet objects for filtering the tweet identifiers: \texttt{created\_at}, \texttt{id}, \texttt{author.verified}, \texttt{author\_id}, \texttt{geo.country}, and \texttt{source}.

\printbibliography[heading=bibliography]
\end{document}